\documentclass[12pt]{article}
\usepackage{eurosym}
\usepackage[T1]{fontenc}
\usepackage[utf8]{inputenc}
\usepackage{geometry}
\usepackage[active]{srcltx}
\usepackage{color}
\usepackage{bbm}
\usepackage{amsthm}
\usepackage{amsmath}
\usepackage{setspace}
\usepackage{lscape}
\usepackage[authoryear]{natbib}
\usepackage[font=footnotesize, width=1\textwidth]{caption}
\usepackage[T1]{fontenc}
\usepackage{graphicx}
\usepackage{times}
\makeatletter

\usepackage{amsfonts}\usepackage[page]{appendix}
\usepackage{color}
\usepackage[usenames,dvipsnames,svgnames,table]{xcolor}
\usepackage{booktabs}
\newcommand{\beq}{\begin{equation}}
\newcommand{\eeq}{\end{equation}}

\begin{document}

\title{Expectations Formation with Fat-tailed Processes: Evidence from Sales Forecasts
\thanks{\scriptsize We thank seminar participants at CFM and Columbia GSB for their constructive feedback. Thesmar is a consultant for CFM.}
\bigskip
} 
\vspace{.5cm}

\author{Eugene Larsen-Hallock\footnote{Columbia \& CFM}
\and Adam Rej \footnote{CFM}
\and David Thesmar \footnote{MIT, CEPR \& NBER}}
\maketitle
\begin{abstract}
We empirically analyze a large sample of firm sales growth expectations. We find that the relationship between forecast errors and lagged revision is non-linear. Forecasters underreact to typical (positive or negative) news about future sales, but overreact to very significant news. To account for this non-linearity, we propose a simple framework, where (1) sales growth dynamics have a fat-tailed high frequency component and (2) forecasters use a simple linear rule. This framework qualitatively fits several additional features of data on sales growth dynamics, forecast errors, and stock returns. 
\end{abstract}

\newpage

\section{Introduction}

Expectations formation is a core question in economics. In recent years, a strain of literature in macroeconomics and finance has been collecting empirical regularities using survey data on subjective forecasts. It finds that forecasts largely deviate from the full information model that predominates in economic modelling: forecast errors are biased and predictable using past errors and past revisions. Two types of explanations for this have been put forward. The first one is that the data-generating process (DGP) is simple and known to forecasters, but forecasting rules are irrational but linear, featuring for instance under-reaction \citep{bouchaud_sticky_2019} or overreaction \citep{bordalo_diagnostic_2019, bordalo_over-reaction_2018, afrouzi_overreaction_2020}. The second approach to explaining observed biases is the tenet that the data-generating process is too complex to be known by forecasters. Thus, they use a misspecified model calibrated on the data they observe. This may come from the fact that the DGP is hard to learn \citep[for recent contributions along these lines see][]{veldkamp, nakamura}, or alternatively from bounded rationality of the forecasters. They can only use simple forecasting rules \citep{fuster_natural_2010, gabaix_behavioral_2018}. In any case, forecast errors are predictable because forecasters use an imperfect model. In this paper, we find evidence consistent with the second view, i.e. that, facing complex (non-Gaussian) processes, forecasters use simple rules.

\medskip

We use data on some 63,601 analyst forecasts of corporate revenue growth and their realizations. An advantage of focusing on revenue growth (instead of EPS as the literature typically does) is that revenue is always positive so that growth rate is always well defined. We first show that the relationship between forecast revisions and future forecast error is non-linear, a feature that is not reported in the existing literature. In some settings, revisions linearly and \emph{positively} predict forecast errors, a feature commonly interpreted as evidence of under-reaction \citep{coibion_information_2015}. In others, revisions linearly and \emph{negatively} predict forecast errors, which is considered as evidence of overreaction \citep{bordalo_diagnostic_2019,bordalo_over-reaction_2018}. In our sample, which is much larger than the existing studies, and which focuses on a rather new object, sales growth, we find evidence of both. For intermediate values of revisions, forecasters underreact to news (an increasing relation between revisions and errors). For large values of revisions, forecaster overreact (a decreasing relation between revisions and errors). This non-linearity is robust. It holds in U.S. data and international data. It holds across most industry groups. 

\medskip

The remainder of the paper is dedicated to explaining this fact. Our framework is based on the simple assumption that forecasters use a linear rule to forecast sales growth, but that this rule is misspecified because the true DGP is more complex. Taking inspiration from the literature on firm size distribution \citep[in particular,][]{axtell, bottazzi_explaining_2006}, we posit that sales growth distribution may be modelled by the sum of a low-frequency and high-frequency shock. The low frequency shock is Gaussian, while the high-frequency shock is non-Gaussian. It may have a very large (positive or negative) realizations. With such a model, the optimal forecast of future growth, conditional on current growth, is non-linear. A perfectly rational forecaster anticipates more reversion to the mean when realizations are extreme and more persistence when realizations are intermediate. We assume, however, that agents stick to a linear rule to make their forecasts. The fact that agents use a misspecified model may be grounded in bounded rationality \citep[i.e., agents use a simple rule even if the process is complex, as in][]{fuster_natural_2010} or the difficulty of learning about complex processes (shocks with multiple frequencies are hard to learn \citealp{nakamura}; shocks with fat tails also \citealp{veldkamp}). 
\medskip

Combined, these two assumptions (linear forecasting rule but short-term non-Gaussian shocks) are enough to generate the non-linear relation between forecast errors and past revisions that observe empirically. The mechanism is intuitive. When revisions are large, the rational forecaster should anticipate mean reversion, but the linear forecaster won't. She overreacts to big positive (or negative) news. When fitting her forecasting rule to the data, she does, however, take this overreaction into account, and optimally attenuates the sensitivity of her forecast to recent observations in the bulk of the distribution. As a result, she underreacts to news of lesser significance.

\medskip

We then qualitatively test four additional predictions of the model. We start with two natural predictions of the data-generating process. The first such prediction is that the distribution of sales growths has fat tails, a fact that holds strongly in the data (and previously shown by \citealp{bottazzi_explaining_2006}). In particular, we check that this fact is not driven by an alternative model of firm dynamics, where firms have heterogeneous volatility, but Gaussian dynamics. In such a setting, large growth shocks could be generated by the subset of firms who are more volatile than average \citep{wyart_statistical_2003}. We thus rescale sales growths by estimates of firm-level standard deviation and find that the resulting distribution still has very fat tails, suggesting that growth shocks occur within firms, not across firms. 

\medskip

The second prediction from our DGP is that, conditional on past growth, future growth should follow a S-shaped pattern as discussed above. We show that this holds in the data, whether we normalize sales growth by firm-level standard deviation or not.

\medskip

The third prediction is on forecast errors. A natural prediction of our forecasting model is that the autocorrelation of forecast errors should have the same non-linear relation as the relation between errors and lagged revision. In our model, where the forecasting rule is linear, they are the same. Large past errors are equivalent to big shocks and therefore transient ones: This leads to overreaction, as in the error-revision relation. We find that this pattern holds in the data: forecast error are positively correlated for intermediate values and negatively for large absolute values.

\medskip

Our fourth and last prediction is on stock returns. Assuming risk-neutral pricing and that equity cash-flows follow a dynamic similar to revenues, it is easy to show that our model predicts that the autocorrelation of returns should have a shape similar to the autocorrelation of forecast errors. For intermediate values of past returns, momentum should dominate, but for extreme values of returns, stock returns should mean revert. We find this pattern to hold in the data. Our findings line up with recent research from \cite{ChristofSchmidhuber}, who also finds evidence of momentum for ``normal past returns'' and mean-reversion for extreme values of returns. We conclude from this analysis that the risk-adjusted performance of momentum strategies would be considerably improved by excluding stocks whose past returns have been large in absolute value.

\medskip

This paper contributes to the recent empirical literature on expectations formation. Most papers in this space focus on linear and Gaussian data-generating processes. Forecasting rules may, or may not, be optimal, but are in general linear, so that the relationship between forecast errors and past revisions (or past errors) is also linear. Our paper emphasizes the non-linearity of such a relation, and as a result, suggests an different modelling approach for the data-generating process to account for this non-linearity. We emphasize non-Gaussian dynamics in firms' growth (as \citealp{veldkamp}, have done in a different setting and in their case with a focus on Bayesian learning). 

\medskip

In doing this we also connect the expectations formation literature with the empirical literature on firm dynamics, which has since \cite{axtell} emphasized the omnipresence of power laws in the distribution of firms sizes (see \citealp{gabaix_power_2009}, for a survey of power laws in economics). That sales \emph{growths} (rather than log sales) have fat tails is a less well-known fact, although it was first uncovered by \cite{bottazzi_explaining_2006}. 

\medskip

Last, our assumption that forecasters use a simple, linear, forecasting rule that is misspecified is inspired by the literature on bounded rationality, which assumes economic agents have a propensity to use oversimplified models to minimize computation costs \citep{fuster_natural_2010, fuster_natural_2012, gabaix_behavioral_2018}. Such models are correct on average, they are fitted on available data, but their misspecification gives rise to predictability in forecast errors. 

\medskip

Section \ref{datasec} describes the data we use: publicly available data on analyst forecasts (IBES) and confidential data on international stock returns from CFM. Section \ref{sec:eform} documents the main fact: future errors are a S-shaped function of past revision. Section \ref{sec:DGP} lays out the simple framework that we build in order to explain this novel pattern. Section \ref{allpredictions} tests four additional predictions from this model. Section \ref{conclu} concludes.

\section{Data}
\label{datasec}

\subsection{Analyst Forecast data}

This paper focuses on firm revenue (sales) forecasts made by analysts. Analyst forecasts come from IBES Adjusted Summary Statistics files, which are available both in the U.S. and internationally. Summary statistics files contains ``current'' estimates as of the third Wednesday of each month. While Earnings per Share forecasts have received greater attention in the literature, sales forecasts are, in fact, better populated in the data than EPS forecasts in recent years. Another advantage of revenue forecasts is that they are never negative, so that we can easily calculate sales growth. A downside of EPS is that it is frequently negative or small rendering the calculation of EPS growth forecast impractical. Thus, the literature on EPS forecasts studies a variable that is, in essence, non-stationary (typically, future EPS normalized by current stock price).

For each firm $i$ and each year $t$, we denote sales by $R_{it}$, and $F_{t}R_{it+1}$ the forecast made in year $t$ for the future realization of $R_{it+1}$. We compute $F_{t} R_{it+1}$ as the consensus three months after the end of fiscal year $t$ (i.e. nine months prior to the end of fiscal year $t+1$) to ensure that sales results for fiscal year $t$ are available when the forecast for $t+1$ is formed. Similarly, the two-year ahead forecast $F_{t-1}R_{it+1}$ is measured three months after the end of fiscal year $t-1$. Finally, we retrieve the realization of $R_{it+1}$ from the IBES actual files, which is designed to recover the realization of the quantity actually forecast by analysts.

\medskip

In this paper, we focus on log sales growth and forecast of log sales growth. We define $g_{it+1}=\log R_{it+1}-\log R_{it}$ the log-growth of this quantity. The growth forecast is defined as $F_t g_{it+1} = \log F_t R_{it+1} - \log R_{it}$ for the one-year ahead growth forecast, and $F_{t-1} g_{it+1} = \log F_{t-1} R_{it+1} - \log F_{t-1} R_{it}$ for the two-year ahead forecast of annual growth. 

\medskip

Finally, in the spirit of the expectations formation literature \citep{coibion_information_2015,bouchaud_sticky_2019}, we construct two empirical variables: the forecast error $ERR_{it+1}=g_{it+1}-F_t g_{it+1}$ and the forecast \emph{revision} $R_t g_{it+1}=F_t g_{it+1}-F_{t-1} g_{it+1}$. These two variables will be the main focus of our analysis. 

\medskip

To ensure forecast quality and improve sample consistency when we examine returns, restrict our analysis of forecasts to firms that belong to one of the major global stock indexes.\footnote{The list of stock markets used consists of: AEX, AS5, CAC, DAX, HSC, HSI, IBE, IND, KOS, MID, NDX, NIF, NKY, OMX, SMI, SPT, RAY, SX5, TOP, TPX, TWY, UKX}. Further, we restrict ourselves to firm-year observations for which both the forecast error $ERR_{it+1}$ and the revision $R_t g_{it+1}$ are available. We give more details about the number of observations and the start date in Table \ref{tab:index_date_exp}. 

\subsection{International Data on Stock Returns}

In examining returns we restrict our sample to equities included in a major national index. We rely on proprietary return data purchased and maintained by CFM. The start of data availability differs by index and is shown in Table \ref{tab:index_date_ret}. For all indexes data has been obtained through January, 2022. Each observation is a ticker-month, and returns are log returns. 

\begin{table}[htbp!]
  \centering
  \caption{Sample size by exchange (sales growth)}
  \label{tab:index_date_exp}
    \begin{tabular}{lrrrrrr}
    \toprule
    Index &  Total &  2000 &  2005 &  2010 &  2015 &  2020 \\
    \midrule
    AEX   &    533 &     0 &    32 &    19 &    30 &    28 \\
    AS5   &   3228 &    48 &   122 &   167 &   196 &   161 \\
    CAC   &    921 &     0 &    40 &    45 &    49 &    48 \\
    DAX   &    680 &     0 &    29 &    38 &    37 &    35 \\
    HSC   &    972 &     7 &    24 &    41 &    75 &    74 \\
    HSI   &    572 &    15 &    24 &    28 &    29 &    29 \\
    IBE   &    715 &     0 &    35 &    37 &    40 &    35 \\
    IND   &    746 &     1 &    38 &    38 &    41 &    39 \\
    KOS   &   1540 &    34 &    29 &    30 &   101 &   124 \\
    MID   &  13016 &    10 &   586 &   818 &   782 &   646 \\
    NDX   &   1174 &     1 &    47 &    67 &    72 &    61 \\
    NIF   &   1037 &    13 &    24 &    47 &    66 &    64 \\
    NKY   &   4959 &   207 &   206 &   226 &   224 &   233 \\
    OMX   &    605 &    19 &    26 &    31 &    32 &    31 \\
    RAY   &  15923 &     4 &   525 &   995 &  1057 &   881 \\
    SMI   &    479 &     8 &    21 &    21 &    27 &    23 \\
    SPT   &    998 &     0 &    34 &    56 &    67 &    63 \\
    SX5   &    215 &     0 &    10 &    11 &    13 &    11 \\
    TOP   &    493 &     0 &    18 &    14 &    40 &    32 \\
    TPX   &  10836 &   372 &   486 &   531 &   504 &   574 \\
    TWY   &   1314 &    13 &    40 &    71 &    80 &    74 \\
    UKX   &   2645 &    82 &   110 &   131 &   142 &   119 \\
    \bottomrule
    \end{tabular}
\end{table}

\section{Motivating Facts}\label{sec:eform}

In this section we describe new evidence on expectations formation and document a strong non-linearity in the link between forecast error and revisions. 


Since \citep{coibion_information_2015-1} many papers in the expectations formation literature estimate the following linear relationship between forecast errors and revision:

\begin{equation}
\label{CG_eq}
ERR_{it+1} = \alpha + \beta R_t g_{it+1} + \epsilon_{it+1}
\end{equation}

\noindent which is intuitive to interpret. Full information rationality predicts $\beta=0$ for consensus forecasts \citep{coibion_information_2015-1}. Plain rationality predicts $\beta=0$ for individual forecasts. $\beta>0$ is typically interpreted as evidence of information frictions \citep{coibion_information_2015}, or, if run at the forecaster level, plain under-reaction (\citealp{bouchaud_sticky_2019}, study EPS forecasts; \citealp{ma_quantitative_2020}, study the revenue forecasts of managers). In contrast, $\beta<0$ is interpreted as evidence of overreaction (\citealp{bordalo_diagnostic_2019}, study long-term EPS growth forecasts; \cite{bordalo_diagnostic_2018-1} focus on macroeconomic expectations). All these papers restrict their analyses to linear functional forms, as in equation (\ref{CG_eq}).

\begin{figure}[htbp!]
    \begin{center}
    \caption{Revenue Forecast Error as a Function of Past Revision}
    \label{fig:zigzag_sales}
    \includegraphics{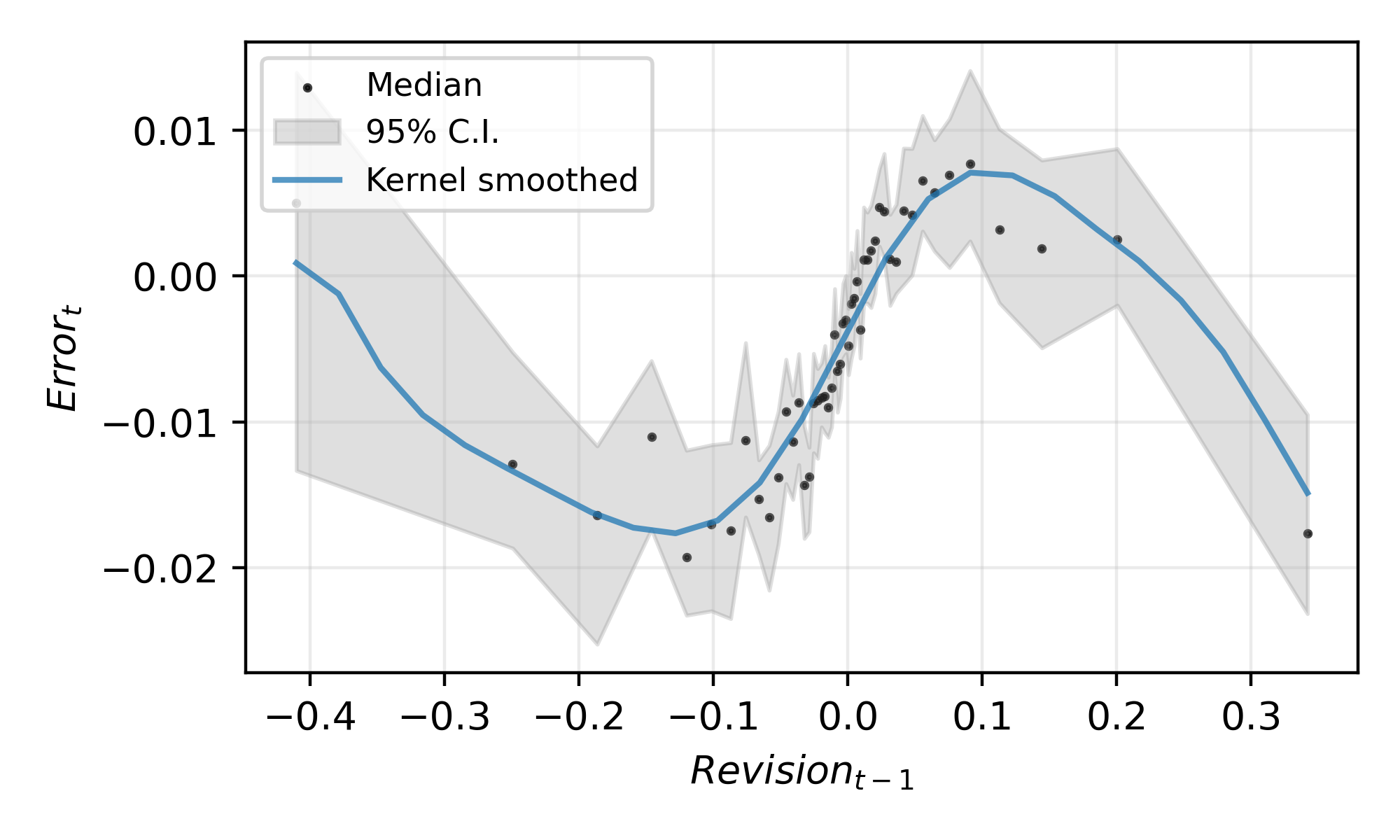}
    \end{center}
    \footnotesize Note: In this figure we use international sample of firm revenue expectations to report the binned scatter plot of future log forecast errors $g_{t+1}-F_t g_{t+1}$ as a function of past revision $F_t g_{t+1} - F_{t-1} g_{t+1}$. The blue line is a local polynomial approximation, centered in the middle of each centile.
\end{figure}

In this section we show that for revenue growth forecasts this relationship is actually non-linear. In Figure \ref{fig:zigzag_sales}, we represent the relationship in a non-parametric way through a binned scatter plot where the x-axis are the revisions $R_t g_{it+1}$ and the y-axis is the average forecast error $FE_{it+1}$. Each black dot corresponds to a centile of the distribution of revisions, with the x coordinate being the average revision in this centile and the y coordinate being the average forecast error. The grey shaded area shows a bootstrapped 95\% confidence interval. The blue line shows the predicted error from a local polynomial regression (or LOESS) model estimated at the center of each percentile of lagged revision. The kernel for this local regression model is Gaussian with the bandwidth set equal to the average of the distances between the centers of the 1st and 2nd, and the 99th and 100th, percentiles. 

\medskip

For revisions of relatively small to moderate magnitude we find that errors are increasing in revision. Thus, forecasters are \textit{under-reacting} in response to moderately-sized news shocks. This consistent with evidence from \cite{bouchaud_sticky_2019} on EPS forecasts in United states. \cite{ma_quantitative_2020} find similar evidence on revenue forecasts from managers' expectations in the U.S. (using guidance data) and Italy (using a survey from the Bank of Italy). Their samples are, however, much smaller than ours (a few 10,000 observations at most), which precludes observing the tails of the distribution of revisions.

The key difference is in the tails of the distribution of revisions, for which this relationship is reversed. In the face of exceptionally bad news, forecasters are \textit{over-reacting}: a larger, positive revision leads to more negative surprises. A similar non-linearity is marginally observable in U.S. EPS forecasts in \cite{bouchaud_sticky_2019}, but the S shape in not complete there.

\medskip

\begin{figure}[htbp!]
    \begin{center}
    \caption{The Error-Revision relationship: Sample Splits}
    \label{fig:zigzag_breakdown}
    \includegraphics[scale=.5]{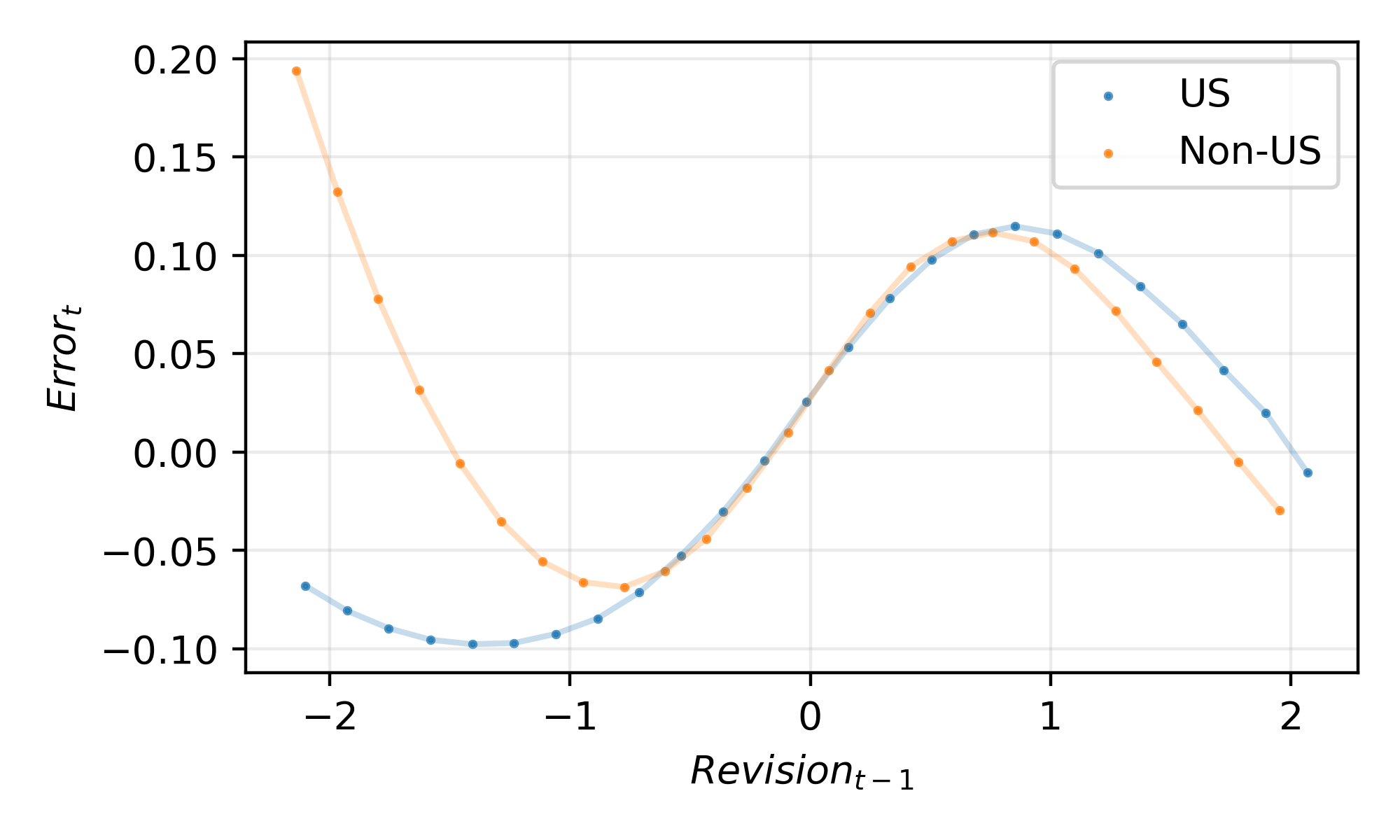}     \includegraphics[scale=.5]{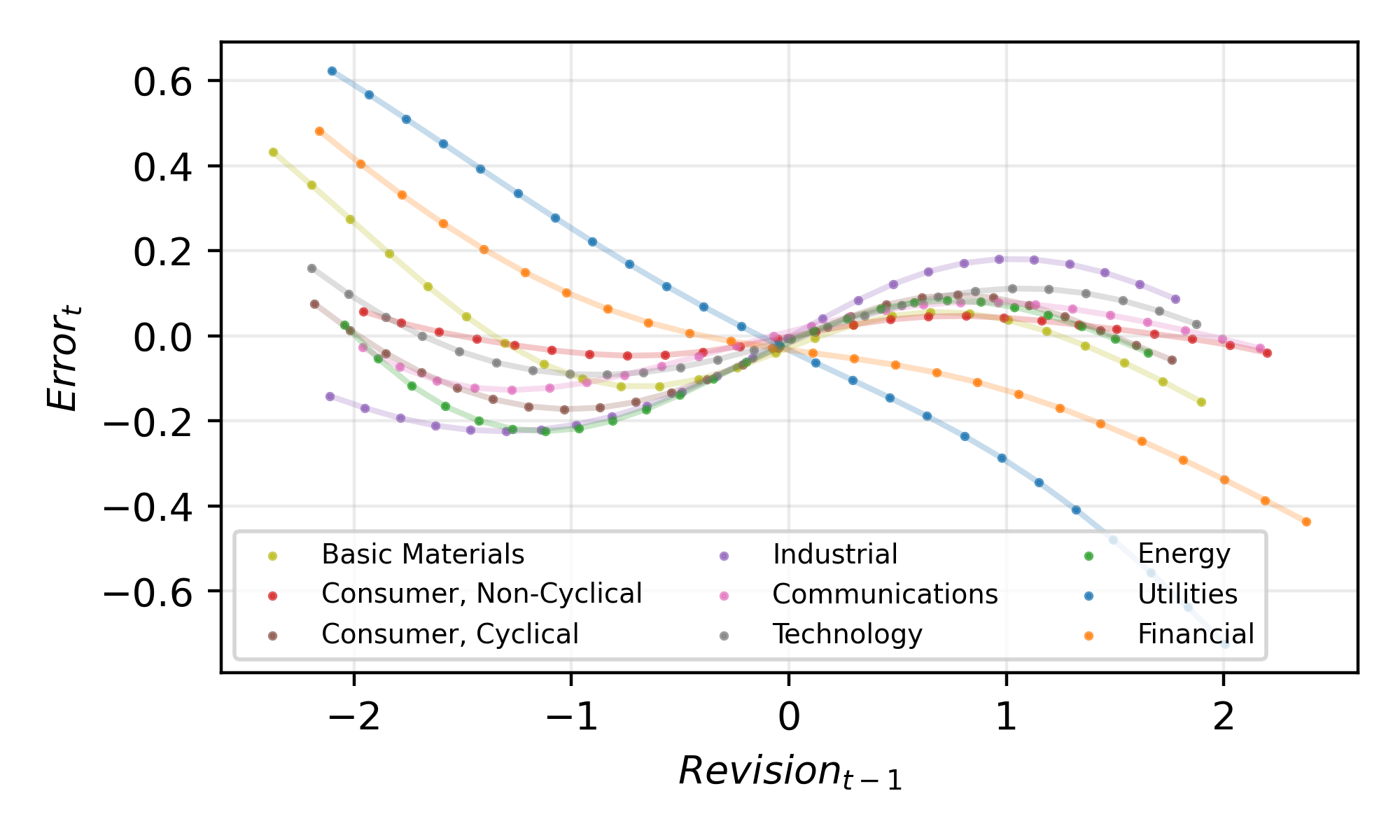} \\
    \vspace{.2cm}
    \footnotesize
    Panel A: US v RoW \hspace{1in} Panel B: By Industry
    \end{center}
    \footnotesize Note: In this figure we use our international sample of firm revenue expectations to report the binned scatter plot of future log forecast errors $g_{t+1}-F_t g_{t+1}$ as a function of past revision $F_t g_{t+1} - F_{t-1} g_{t+1}$. The blue line is a local polynomial approximation, centered in the middle of each centile. In Panel A, we split the sample between U.S. and non U.S. observations. In Panel B, we split the sample into 1 digit GICS industries.
\end{figure}

We then explore the robustness of this relationship across sub-categories in Figure \ref{fig:zigzag_breakdown}. This figure has two panels: one that splits between U.S. and non-U.S. firms (Panel A) and one that splits the sample into industries (Panel B). In both cases we only show the prediction from flexible polynomial approximation. In both subcategories the S-shaped function emerges. In particular, it is visible both in US and non-US firms, although more pronounced among U.S. firms. 

Overall, the evidence on log forecast errors and revisions points towards a different treatment of large v. smaller shocks. Such evidence is hard to square with established models of expectations formations, which feature linear DGPs (typically, AR1 models) and linear expectations models. In what follows, we set up a simple model that features extreme (i.e. non-Gaussian) shocks and linear expectations formation in order to captures the above non-linearity.

\section{Model}

In this section we develop a parsimonious model that features extreme shocks and linear expectations in order to capture the non-linear behavior of expectation errors of Figure \ref{fig:zigzag_sales}.

\subsection{Modeling Sales Growth}
\label{sec:DGP}

The first piece of the model is the data-generating process. We will omit the firm index $i$ for clarity's sake and assume that log sales growth, $g_{it+1}$, evolves according to:

\begin{align}
    \label{g}
    g_{t+1} &= \underline{g}_{t+1} + \epsilon_{t+1} \\
    \label{mu}
    \underline{g}_{t+1} &= \overline{\underline{g}} + \phi(\underline{g}_t - \overline{\underline{g}}) + u_{t+1} 
\end{align}

\noindent where $\underline{g}_{t+1}$ is the unobservable latent state that follows the classic linear-Gaussian AR1 dynamics. The key difference with most existing models of expectations formation is that $\epsilon_{t+1}$ follows a probability distribution density that has heavy tails. Because it fits the data quite well (as we document below), we assume that $\epsilon_{t+1}$ follows a Student's t distribution with $\nu$ degrees of freedom. Thus:

\begin{align*}
    \epsilon_{t+1} &\sim \text{Student-t}(0,1, \nu) \\
    u_{t+1} &\sim \text{Normal}(0,1) 
\end{align*}

Note that, although we analyze the cross-section of firms, we assume a single process for all firms. In this paper, we do not explore the consequences of firm heterogeneity for forecasting biases. For instance, such biases could arise from forecasters using one single forecasting model for firms following different processes. We believe such an avenue is interesting, but beyond the scope of this paper, which focuses on one single deviation from the classical model, i.e. that temporary shocks have fat tails. In order to bring the data closer to the model however, we will conduct all of our analysis with ``normalized growth'' data, thereby ensuring that all firms have the same growth volatility. We discuss this adjustment extensively in Section \ref{allpredictions}.

\medskip

\begin{figure}[htbp!]
    \caption{Actual vs. lagged actual in model simulation.}
    \label{fig:act_sim}
    \begin{center}
    \includegraphics{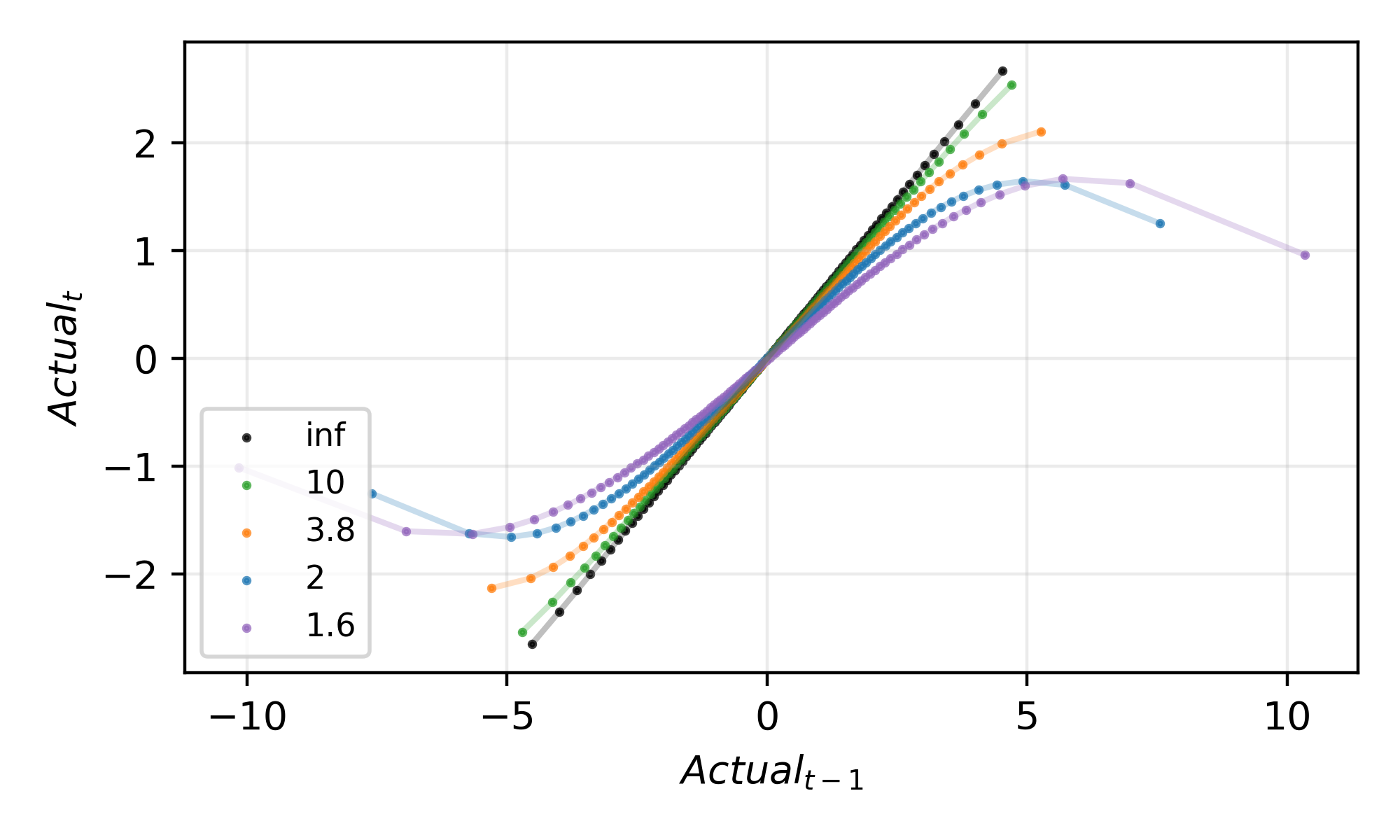}
    \end{center}
    \footnotesize Note: We simulate the model over T periods with $u$ following a normal distribution $N(0,1)$, and $\epsilon$ following  $\text{Student-t}(0,1, \nu)$. We then show local polynomial regressions of $g_t$ on $g_{t-1}$ estimated at the center of each percentile of lagged realization $g_{t-1}$. We explore values of $\nu$ from 1.6 (fat tailed) to $\infty$ (Gaussian). 
\end{figure}

In our simple model the conditional expectation $E\left(g_{it+1} | g_{it}\right)$ is non-linear. We show this numerically in Figure \ref{fig:act_sim}. For different values of $\nu$, we numerically simulate the process and compute the conditional expectation $E\left(g_{it+1} | g_{it}\right)$ on simulated data. As shown in Figure \ref{fig:act_sim} this relationship is indeed quite linear in the body of the distribution, but experiences ``reversals'' in the tails. While not visible in Figure \ref{fig:act_sim}, all finite values of $\nu$ leads to such reversals in the tail, but as $\nu$ gets larger (and $\epsilon$ is closer to being Gaussian) they get pushed out farther into the tails and are very sharp and localized.

The economic intuition is simple. Since the underlying state variable is Gaussian, extreme negative or positive realizations are more likely to come from the transitory process $\epsilon$ than the persistent one $\underline{g}$, since it features more extreme shocks. As a result, a large sales growth realization today is unlikely to translate into future large sales growth tomorrow: This suggests the presence of ``reversals'' in the tails, as we see in Figure \ref{fig:act_sim}.

The above process has several predictions about the distribution of growth rates, one of them being that the cross-sectional distribution of growth rates should have fat tails. We will explore these predictions in Section \ref{allpredictions}.

\subsection{Expectations Formation}
\label{sec:expfor}

The second building block of the model is the  formation of expectations. Our core assumption is that forecasters fail to perceive the non-linearity of true expectations $E\left(g_{it+1} | g_{it}\right)$ and use a linear rule. This assumption is based on the idea that economic agents use simplified, ``sparse'', model of reality to formulate expectations \citep{fuster_natural_2012,gabaix_behavioral_2018}. Agents assume $g_{it}$ follows a linear AR(p) model, estimate it on data and use this model to form forecasts. One advantage of this representation is that the term structure of forecasts is naturally defined, as agents calculate mathematical expectations under the AR(p) model. Hence, we assume that the forecaster believes growth follows the following AR(p) model:

$$g_{t+1} = \underline{g} + \sum_{s=0}^{p-1}{\beta_k \left(g_{t-k}-\underline{g}\right)} + u_{t+1}$$

\noindent We denote the subjective expectation operator by $F_t g_{t+k} \equiv EL(g_{t+k}|\underline{g_{t}})$.



We assume that this prior is dogmatic. The forecaster is willing to re-estimate the model's parameters as new data comes in, but does not explore models outside of the AR(p) set-up. As a result, the agent does not really formulate rational expectations since she does not estimate the right DGP, as in \cite{fuster_natural_2010}. One foundation for such dogmatism is that learning is extremely slow in non-Gaussian, non-linear environments, so that it takes many periods to modify the prior about the model (in recent literature, see \cite{veldkamp} and \cite{nakamura}). 

Thus the agent estimates the parameters of the misspecified model using OLS on expanding windows -- using all information until date $t$. Let $\widehat{\underline{g}}$ and $\widehat{\beta_k}$ be these estimates. The one-period ahead forecast and the revision are given by:

\begin{align*}
    F_t g_{t+1} & = \widehat{\underline{g}} + \sum_{s=0}^{p-1}{\widehat{\beta_k} \left(g_{t-k}-\widehat{\underline{g}}\right)}
\end{align*}

\subsection{Predictions of the Model: Errors on Revisions}

We now check that our model indeed generates the non-linear relation between revenue forecast errors and revenue forecast revisions shown in Figure \ref{fig:zigzag_sales}. 

\begin{figure}[htbp!]
\begin{center}
    \caption{Error as a function of lagged revision}
    \label{fig:sim_err_l_rev}
    \includegraphics{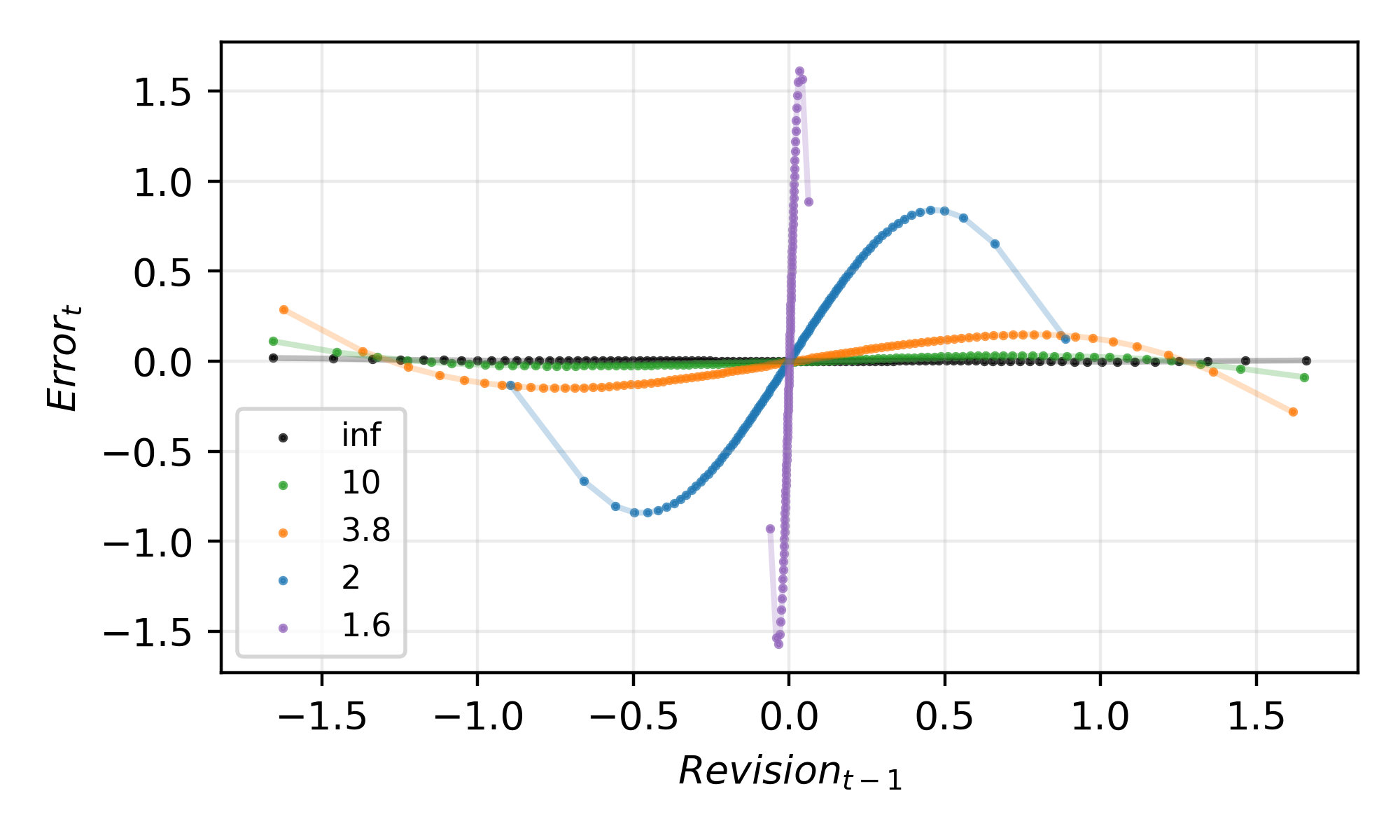}  
\end{center}    
\footnotesize Note: We simulate the model over T periods with $u$ following a normal distribution $N(0,1)$, and $\epsilon$ following  $\text{Student-t}(0,1, \nu)$. We then show local polynomial regressions of error $ERR_{t+1}=g_{t+1} - F_t g_{t+1}$ on revision $R_t g_{t+1}= F_t g_{t+1} - F_{t-1} g_{t+1}$ estimated at the center of each percentile of revision. We explore values of $\nu$ from 1.6 (fat tailed) to $\infty$ (Gaussian). Forecasters are assumed to employee an AR(3) model when predicting dividend growth rates.
\end{figure}

In Figure \ref{fig:sim_err_l_rev}, we report results from simulations, assuming that forecasts are based on a fitted AR(2) model. We vary the thickness of the tail of the temporary shock $\epsilon$, which is governed by $\nu$. $\nu=+\infty$ corresponds to a normal distribution, while $\nu=1.6$ is the thickest tail possible. 

Figure \ref{fig:sim_err_l_rev} shows that, as long as the temporary shock has sufficiently fat tails, the linear expectations model generates predictable forecast error that display a non linear pattern similar to Figure \ref{fig:zigzag_sales}. This is quite intuitive. As shown previously, the true conditional expectation is non-linear (see Figure \ref{fig:act_sim}). When the past realization of revenue growth is large, it is likely that it was driven by the temporary fat-tailed process. As a result, the rational forecaster would expect some mean-reversion, but the linear forecaster does not. This creates overreaction to large shocks. In contrast, when past realizations are moderate, there is underreaction. This comes from the fact that the linear forecaster is on average rational: She fits a linear relation on the S-shaped data of Figure \ref{fig:act_sim}. The slope of forecasts for smaller realizations incorporates some of the overreaction in the tails. 

To gain further insights in Figure \ref{fig:act_sim} we vary the fatness of the tail $\nu$. The less thick-tailed the innovation process, the less predictable errors are. When $\nu=+\infty$, the temporary shock $\epsilon$ is Gaussian and forecast errors are very close to zero for all lagged realizations (the black dots line up on the x-axis). This is because in this case the linear AR2 forecasting rule is nearly rational. Indeed, in this case, the rational expectation is a Kalman filter: 

$$K_t g_{t+h} =\underline{g} +  \phi^h G\sum_{s=0}^{+\infty}{\left(1-G\right)^s \left(g_{t-s}-\underline{g}\right)}$$

\noindent where $G$ is the Kalman ``gain''. The AR2 process is close enough to the above equation that forecast errors are nearly zero in our simulations.

The bottom line of this analysis is that the non-linear structure of expectations errors can easily arise when forecasters use linear models when the data generating process has temporary shocks that have fat tails. Indeed in this case, the optimal forecasting rule is non-linear, even though the process is itself linear.


\subsection{An Additional Prediction: Error on Lagged Error}

The empirical expectations literature also investigates a different moment: The autocorrelation of expectation errors (for instance, \cite{ma_quantitative_2020} and \cite{nakamura} among many others).

In our model the autocorrelation of errors is equivalent to the error-revision coefficient. This happens because revisions are directly proportional to current forecast errors:

\begin{equation}
\label{ERRREVlin}
\underbrace{F_t g_{t+1}- F_{t-1} g_{t+1}}_{\equiv R_t g_{t+1}} = \widehat{\beta_0}\cdot\left(g_t-F_{t-1} g_{t}\right) 
\end{equation}

\noindent which means that a positive surprise translates into a positive revision about future growth. The fact that the prior is linear makes this relationship linear, whatever the number of lags $p$. 

As a result, we expect the non-linear relation between errors and lagged revision of Figure \ref{fig:zigzag_sales} to also hold between error and lagged \emph{errors}. We test this additional prediction in Section \ref{allpredictions}.

\subsection{Model Prediction on Returns: Building Intuition}
\label{returns_model}

We also derive predictions on stock returns. Our simple model, as we will see, predicts that momentum occurs for intermediate returns and mean-reversion occurs for extreme returns. 

\medskip

In the spirit of \cite{bouchaud_sticky_2019}, we assume stock prices are given by:

\begin{equation}
\label{APequation}
P_t = \sum_{s \geq 1 }{\frac{F_t D_{t+s}}{\left(1+r\right)^s}}
\end{equation}

\noindent where $F_t D_{t+s}$ is based on the forecasting rule described above in Section \ref{sec:expfor}. Hence, the stock is priced by investors who form expectations based on a linear AR2 fitted on past realizations, while we assume dividends to follow the process described in Section \ref{sec:DGP}. We also assume for simplicity that investors are risk-neutral, so that the discount rate is fixed at $r$. 

In this very simple asset pricing model we expect returns to be a non-linear function of past returns, similar to what we documented for the error-revision relation in Figure \ref{fig:zigzag_sales}. Before we discuss simulation results and economic intuition, it is worth showing the algebra. The standard first order Campbell-Shiller approximation writes as:

$$ r_{t+1} - F_t r_{t+1} \approx \left( F_{t+1}-F_{t} \right) \sum_{s=0}^\infty \rho^s  g_{t+1+s} - \left( F_{t+1}-F_{t} \right)  \sum_{s=1}^\infty \rho^s  r_{t+1+s}$$

\noindent where we denote log dividend growth as $g$ with a slight abuse of notation ($g$ stands for log revenue growth in the rest of the paper). Equation (\ref{APequation}) assumes constant expected returns $ F_t r_{t+k} = r$ (investors may be biased but are risk neutral), so that the CS decomposition simplifies into:

\begin{align*}
  r_{t+1} - r &= \left( F_{t+1}-F_{t} \right) \sum_{s=0}^\infty \rho^s  g_{t+1+s} \\
              &= g_{t+1} - F_t g_{t+1} + \rho \left( F_{t+1}-F_{t} \right) \sum_{s=0}^\infty \rho^s g_{t+2+s} \\
\end{align*}

It then remains to describe the infinite sum of discounted dividend growth. 
In this paper, we assume that forecasters (mistakenly) estimate dividend growth as an AR(p) process:

$$ g_t - \underline{g} = \sum_{s=1}^p \beta_s \left(g_{t-s} - \underline{g} \right) + \epsilon_t$$

We can then stack the estimated AR(p) coefficients $\beta_1,\dots,\beta_p$ into ``companion'' form:

$$ \begin{bmatrix}
g_{t} - \underline{g} \\
g_{t-1} - \underline{g} \\
\vdots \\
g_{t-p} - \underline{g}
\end{bmatrix} 
= \begin{bmatrix}
\beta_1 & \cdots & \beta_{p-1} & \beta_p \\
1 & \cdots & 0 & 0 \\
\vdots & \ddots & 0 & 0 \\
0 & \cdots & 1 & 0  
\end{bmatrix}
\begin{bmatrix}
g_{t-1} - \underline{g} \\
g_{t-2} - \underline{g} \\
\vdots \\
g_{t-p-1} - \underline{g}
\end{bmatrix} 
+ \begin{bmatrix}
\epsilon_t \\
0 \\
\vdots \\
0
\end{bmatrix} $$

\noindent or more compactly:

$$ \mathcal{G}_t = \mathbf{B}\mathcal{G}_{t-1} + \mathbf{\epsilon}_t $$

Time $t$ forecasts for $g_{t+s}-\underline{g}$ are then given by:

$$F_t (g_{t+s}-\underline{g}) = \mathbf{e}_1' \mathbf{B}^{s}\mathcal{G}_{t} $$

\noindent where $\mathbf{e}_1$ is a ``selector'' vector picking out the first element of the following vector. The infinite sum of discounted forecast dividend growth is given by:

\begin{align*}
    F_t \sum_{s=0}^\infty \rho^s (g_{t+1+s}-\underline{g}) &= \sum_{s=0}^\infty \rho^s \mathbf{e}_1' \mathbf{B}^{s+1}\mathcal{G}_{t} \\
    &= \mathbf{e}_1' \mathbf{B}(\mathbf{I}-\mathbf{B})^{-1}\mathcal{G}_{t}
\end{align*} 

We then plug this formula into the CS decomposition and obtain: 

\begin{align}
    r_{t+1} - r &= \mathbf{e}_1' \left( \mathcal{G}_{t+1} - \mathbf{B}\mathcal{G}_{t} \right) + \rho\mathbf{e}_1'\mathbf{B}(\mathbf{I}-\mathbf{B})^{-1} \left( \mathcal{G}_{t+1} - \mathbf{B}\mathcal{G}_{t} \right) \\
                &= \mathbf{e}_1' \left(\mathbf{I} + \rho\mathbf{B}(\mathbf{I}-\mathbf{B})^{-1} \right) \underbrace{\left( \mathcal{G}_{t+1} - \mathbf{B}\mathcal{G}_{t} \right)}_{=ERR_{t+1}\mathcal{G}_{t+1}}
\end{align}

The above expression shows that under the AR(p) assumption, returns are a linear function of past forecast errors ($ERR_{t+1}\mathcal{G}_{t+1}$ is the vector of past $p$ forecast error). In this simple asset-pricing model, returns are only predictable if dividend growth forecast errors are predictable. Under rational expectations (i.e. if the true DGP for dividends is an AR(p)), they are not. But if dividends are driven by a thick-tailed state variable, the true DGP is far from AR(p) as we have documented. Thus, expected forecast errors are non-linear functions of past errors, and the same should hold for returns and past returns. So our model predicts that returns should be a non-linear function of past returns, in other words, momentum should only be present for intermediate values of past returns.

\subsection{Model Prediction on Returns: Simulations}

\begin{figure}[htbp!]
    \begin{center}
    \caption{Momentum for Intermediate Past Returns; Reversal in the Tails}
    \label{fig:sim_ret_l_ret}
    \includegraphics{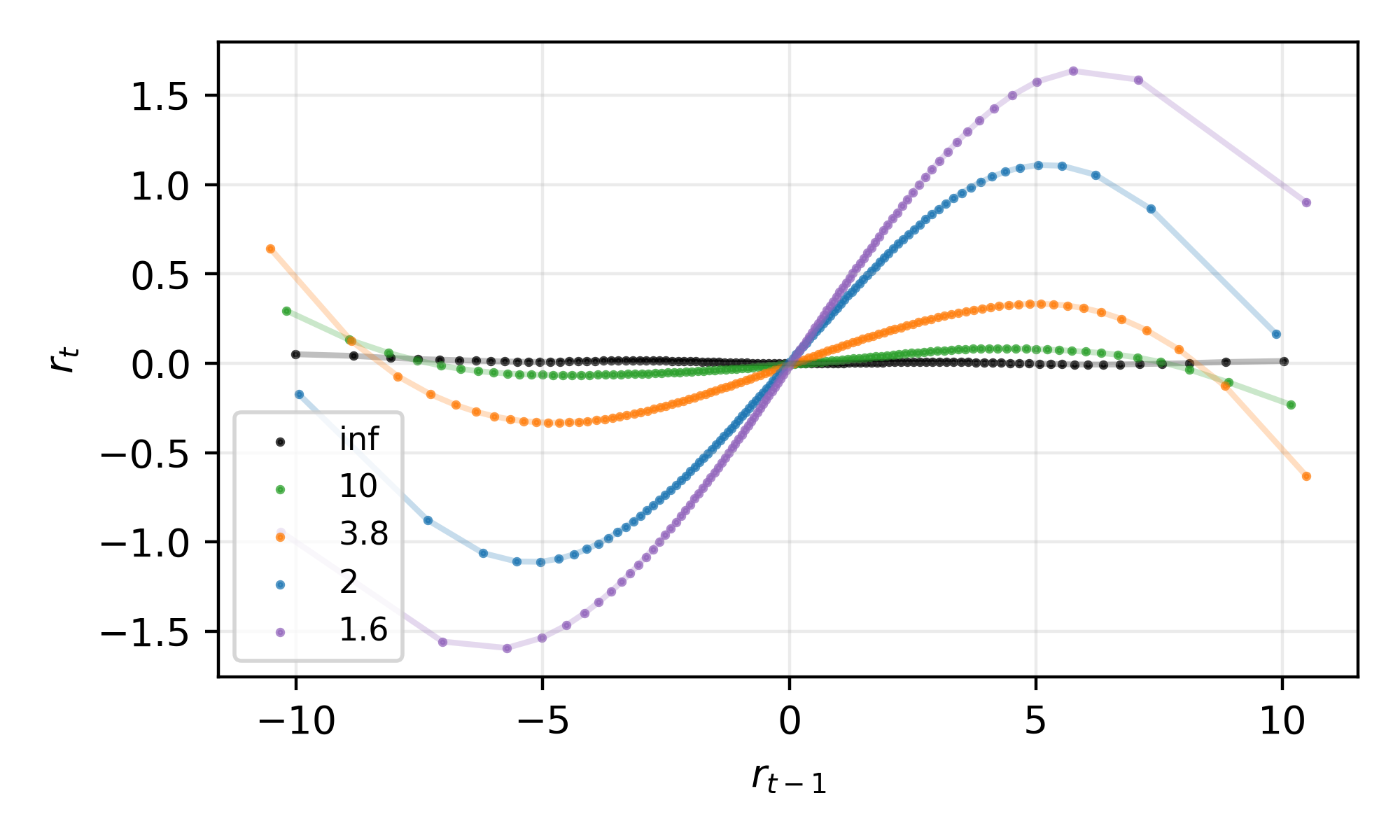}
    \end{center}
    \footnotesize Note: We simulate the model over many periods with $u$ following a normal distribution $N(0,1)$, and $\epsilon$ following  $\text{Student-t}(0,1, \nu)$. We then show local polynomial regressions of $R_{t+1}$ on $R_{t}$ estimated at the center of each percentile of lagged realization $R_t g_{t+1}$. We explore values of $\nu$ from 1.6 (fat tailed) to $\infty$ (Gaussian). Forecasters are assumed to employee an AR(3) model when predicting dividend growth rates. 
\end{figure}

In order to check that this prediction also holds without CS approximation, we proceed to simulate our model. On simulated data, we build returns as $R_{t+1}=(P_{t+1}+D_t-P_t)/P_t$. We then plot average future returns by bins of past returns in Figure \ref{fig:sim_ret_l_ret}. In this very simple asset pricing model, returns are predictable as soon as the dividend process 

The intuition is the same one as before. Very high past returns likely emerge from surprises due to a large thick-tailed temporary shock. Linear forecasters overreact: large past return events are likely to be situations where dividend realization a one-off boon. As a result, linear forecasters overestimate the level of future dividends: The stock price rises too much and future returns are lower. Intermediate past returns, however, are likely generated by standard shocks. There, the linear forecaster under-reacts to small dividend news and the price does not respond enough to these news. Future returns are thus positively correlated with past returns for ``smaller'' absolute values of returns. Note that when $\nu=+\infty$, temporary shocks are Gaussian and, as discussed previously, linear expectations are quasi optimal -- a true Kalman filter would be perfectly rational -- and past returns do \emph{not} predict future returns. 

\section{Testing the Model's Predictions}
\label{allpredictions}

\subsection{Predictions of Growth Rate Dynamics}

In this section we discuss two key predictions of our data-generating process (\ref{g})-(\ref{mu}). The first one is that the distribution of firms \emph{revenue growths} should have fat tails. The second one is that the conditional expectation of $g_{t+1}$ on $g_t$ should be non-linear as in Figure \ref{fig:act_sim}.

\medskip

First, a key prediction of our DGP (\ref{g})-(\ref{mu}) is that the distribution of firm size growth has fat tails. Many variables relevant for finance and economics are not normally distributed \cite{gabaix_power_2009}. It is for instance well-known that the distribution of firm sizes follows a Zipf law \citep{axtell}. Less well-known is also the fact that the distribution of firm \emph{growth rates} has fat tails. \cite{bottazzi_explaining_2006} show that the distribution of Compustat firms follows a Laplace distribution. Here, we provide similar evidence from our sample.

\begin{figure}[htbp!]
    \begin{center}
    \caption{Tail of log sales growth distribution: Fit of various distributions}
    \label{fig:qq_sal}
    \includegraphics{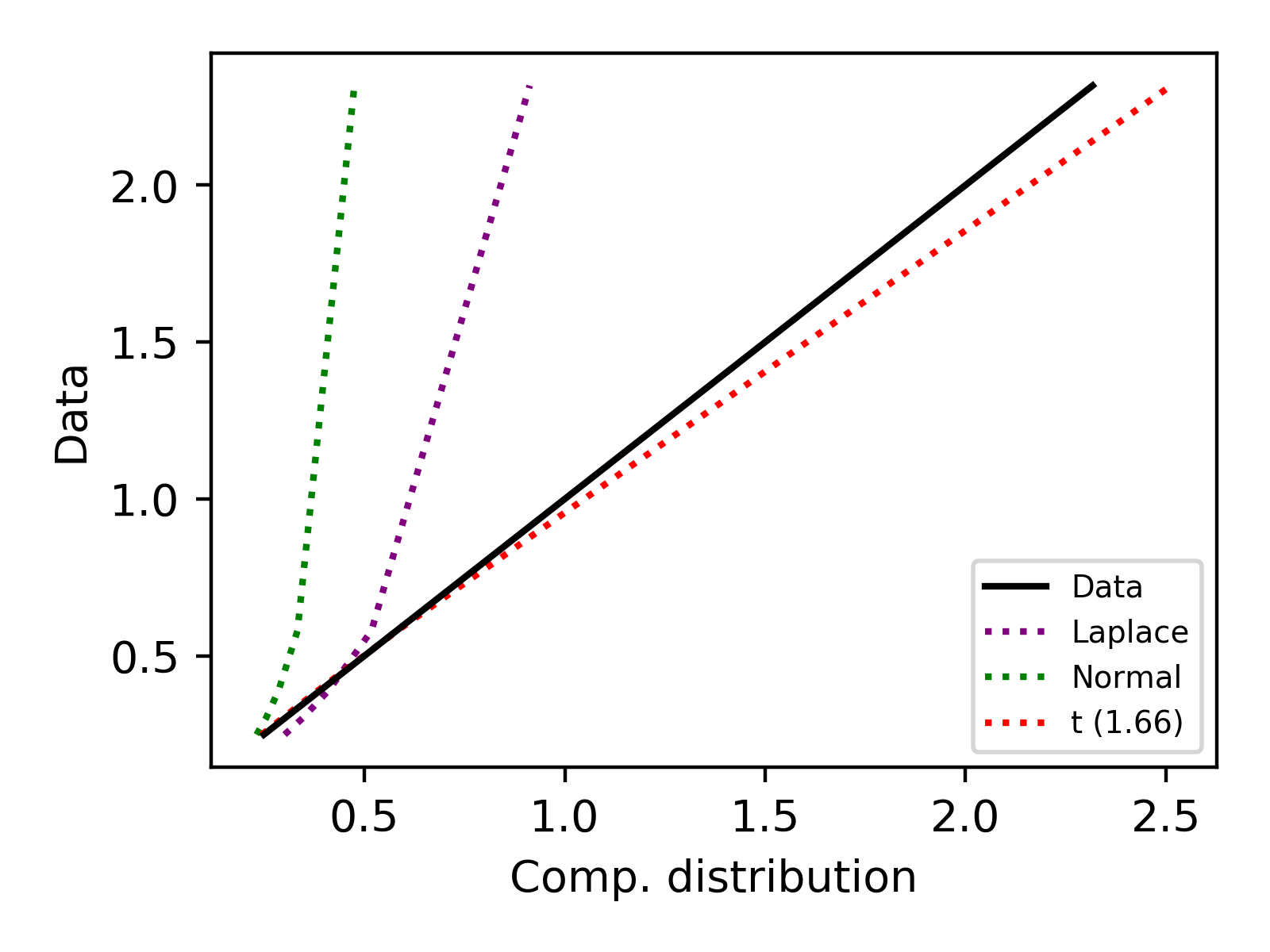}
    \end{center}
    \footnotesize Note: This is a Q-Q plot of log sales growth vs. some textbook distributions (Laplace, Normal and Student). This plot shows the tail above the 90\textsuperscript{th}-percentile of the distribution of $|g_{it}|$. On the x-axis, we report the value of the quantile ($F^{-1}(q)$) of the comparison distribution. On the y-axis, we report the value of the same quantile of the data distribution ($F_{\text{data}}^{-1}(q)$). The data distribution is normalized so that its variance is one. The comparison distribution also have unit variance. By design, the ``data'' line is the 45 degree line.
\end{figure}

In Figure \ref{fig:qq_sal} we show the QQ plot of the sales growth distribution in our sample, along some other textbook distributions. This QQ plot focuses on observations above the 90$^{th}$ percentile distribution of $|g_{it}|$, the \emph{absolute} value of sales growth (so both negative and positive shocks). Each point of this chart corresponds to one quantile of the data distribution. For a given quantile $q$ and a c.d.f. $F$, the y coordinate of the point is the average value of absolute growth at quantile $q$ of the data distribution. Since we focus on the top 10\% of absolute sales growth, this number is positive (so the y axis does not start at zero). The x coordinate of that point is the average value of the \emph{same quantile} of the chosen distribution $F$, or $F^{-1}(q)$. The closer is $F$ to the data distribution, the more the chart will look like a 45 degree line (the black line on the Figure). 

Looking at Figure \ref{fig:qq_sal} it is clear that the distribution of growth rates is very different from normal in the tail. The green line increases faster than the 45 degree line meaning that the sales growth distribution has much heavier tails than normal distribution. The best fit is obtained by fitting a Student distribution.

\medskip

Our model crucially assumes that this distribution comes from temporary shocks occurring \emph{within} firms. \cite{wyart_statistical_2003} suggest an alternative explanation for such thick tails: sales growth has a normal distribution at the firm-level, but that the standard deviation of this process varies across firms. In this case, extreme growth rates would typically occur among firms that have very volatile growth rates (for instance, smaller firms). This alternative interpretation does not explain our findings, but it is worthwhile to analyze its validity.

\medskip

In order to do that we normalize growth rates by a measure of firm-level ``volatility''. To do this we compute the mean absolute deviation of log sales for each firm. This measure of volatility has the advantage of being more immune to fat tails in the growth distribution (since variance may not exist in such cases). For firm $i$, we thus compute:

$$MAD_i=\frac{1}{T_i}\sum_{t=0}^{T_i} |g_{it}-\overline{X}_i|$$

\noindent where $T_i$ is the number of observations for the firm, and $\overline{X}_i$ is the average sales growth at the firm level. 

\medskip

\begin{figure}[htbp!]
    \begin{center}
    \caption{Tail of log sales growth distribution (Normalized)}
    \label{fig:qq_sal_norm}
    \includegraphics{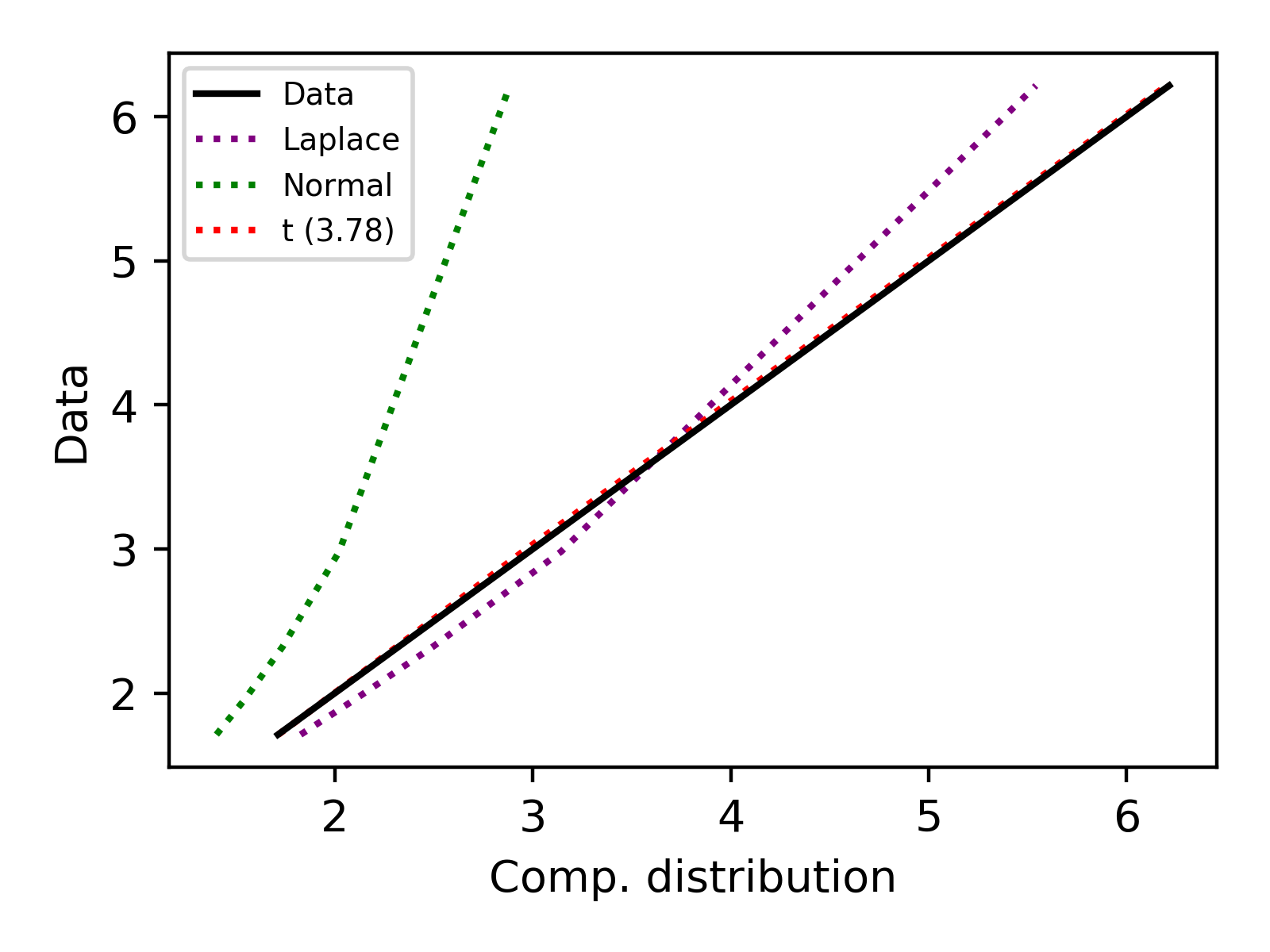}
    \end{center}
    \footnotesize Note: This is a Q-Q plot of \emph{normalized} log sales growth vs. some textbook distributions (Laplace, Normal and Student). This plot shows the tail above the 90\textsuperscript{th}-percentile of the distribution of $|g_{it}-\overline{g}_i|/MAD_i$. On the x-axis, we report the value of the quantile ($F^{-1}(q)$) of the comparison distribution. On the y-axis, we report the value of the same quantile of the data distribution ($F_{\text{data}}^{-1}(q)$). The data distribution is normalized so that its variance is one. The comparison distribution also have unit variance. By design, the ``data'' line is the 45 degree line.
\end{figure}

In Figure \ref{fig:qq_sal_norm} we show the QQ plot of the distribution of normalized $\frac{g_{it}}{MAD_i}$. If heavy tails are driven by firms with larger growth variance, this adjustment should significantly reduce the fat tails of the data. The QQ plot shows that the distribution is still strongly non-normal, though now the fit of the Laplace distribution is much better (consistent with \citealp{bottazzi_explaining_2006}) and the one of the student distribution is nearly perfect.

From this analysis we draw the conclusion that sales growth distribution has heavy tails and that the conditional expectation $E\left(g_{it+1} | g_{it}\right)$ is non-linear. We will postulate below a model of growth dynamics that fits these two facts, and show how it can explain the non-linear relation between revisions and errors.

\bigskip

From now on we report results using the above normalization by the mean absolute distance. This allows to account for fat-tails effects stemming from heterogeneous growth variance. Also quite importantly our model (\ref{g})-(\ref{mu}) assumes homoskedasticity, so it is important to rescale the data so that they have the same property.

\bigskip

A second key prediction of our Data is that the conditional expectation $E_t \left( g_{t+1} | g_{t} \right)$ should be non-linear, as shown in Figure \ref{fig:act_sim}. As mentioned previously the intuition is that large revision presumably come from large shocks to firm sales growth. Since in our model large shock are transitory, the rational forecaster should not expect that large shocks should persist going forward. Smaller shocks are, however, much more likely to stem from the permanent component of revenue, and therefore the rational forecaster should expect them to persist. We now check whether this relationship holds in the data.

\begin{figure}[htbp!]
    \begin{center}
    \caption{Future Growth as a Function of Past Growth (Normalized)}
    \label{fig:act_data}
    \includegraphics{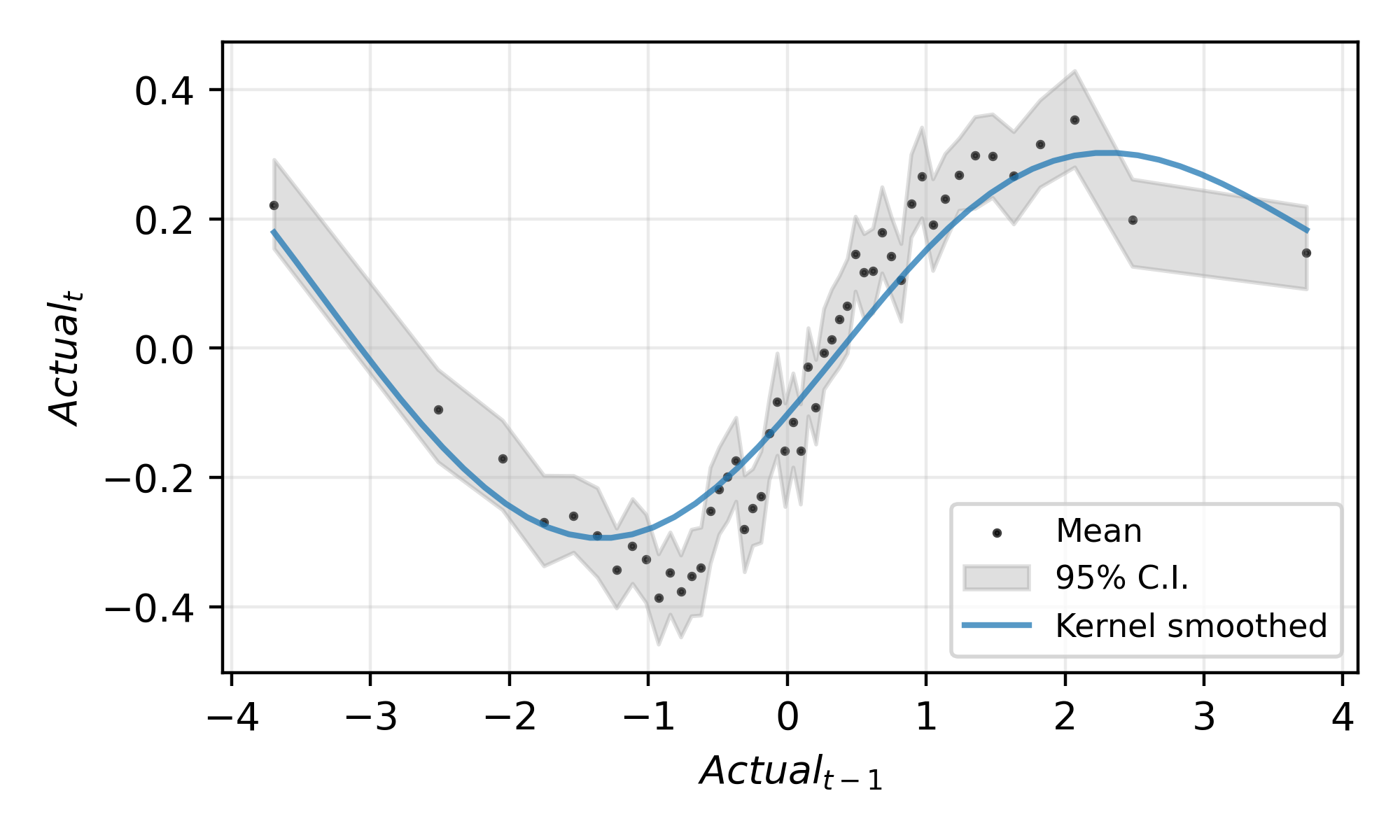}
    \end{center}
    \footnotesize Note: This Figure reports the binned scatter plot of future growth $|g_{it+1}-\overline{g}_i|/MAD_i$ by value of past growth $|g_{it}-\overline{g}_i|/MAD_i$. Each bin corresponds to a centile of the past growth distribution. The blue line is a local polynomial approximation centered around each one of these centiles.
\end{figure}

In Figure \ref{fig:act_data} we construct a binned scatter plot of sales growth against lagged sales growth. To make sure all firms have the same growth volatility (as in the model), we normalize growth by our estimate of the firm level standard deviation $MAD_i$. Each black dot on this figure represents a centile of the distribution of lagged log growth of sales. The x axis shows the average lagged growth and the y axis measures the average current growth. This chart shows that the relationship between current and lagged growth is far from being linear and looks like the S curve shown in Figure \ref{fig:zigzag_sales}. For intermediate levels of growth (between 0 and 1 standard deviation), past growth translates into higher future growth, with a coefficient of about 0.3. The relationship does, however, become much flatter for high growth (with a slightly negative slope in the tail). For negative growth the slope becomes strongly negative. The lower the past growth, the higher the future growth will be, which is consistent with the idea of a rebound. Conditional on survival, very poor past performance predicts strong future growth, as in our model.

\subsection{Predicting Forecast Errors}

The model was designed to predict that forecast errors be a non-linear function of past revisions. Given that the model assumes that all firms have the same variance of shocks, it is natural to check that our main empirical results holds after rescaling by firm-level variance. Another reason why it is important to perform such a robustness check is discussed is discussed in the previous Section. Assume, following \cite{wyart_statistical_2003}, that large news purely come from a separate group of firms (those with more volatile, but still Gaussian, shocks). Then, if forecasters use a firm-level linear forecasting rule, then their forecast errors should be close to unpredictable (to the extent that the AR(p) model they use mimics the optimal Kalman filter).\footnote{If, however, forecasters were to use a global forecasting rule (a single rule estimated on all firms), the non-linear shape may be predicted. Indeed, assume all shocks are Gaussian, but firms differ in the volatility of their \emph{temporary} shock $\epsilon$. In this case, a unique forecasting rule would overestimate the persistence of large shocks. We do not explore this lead in this paper since we have documented in the previous Section that normalized growth is far from being normally distributed.}

\begin{figure}[htbp!]
    \begin{center}
    \caption{Revenue Forecast Error as a Function of Past Revision (Normalized)}
    \label{fig:zigzag_sales_norm}
    \includegraphics{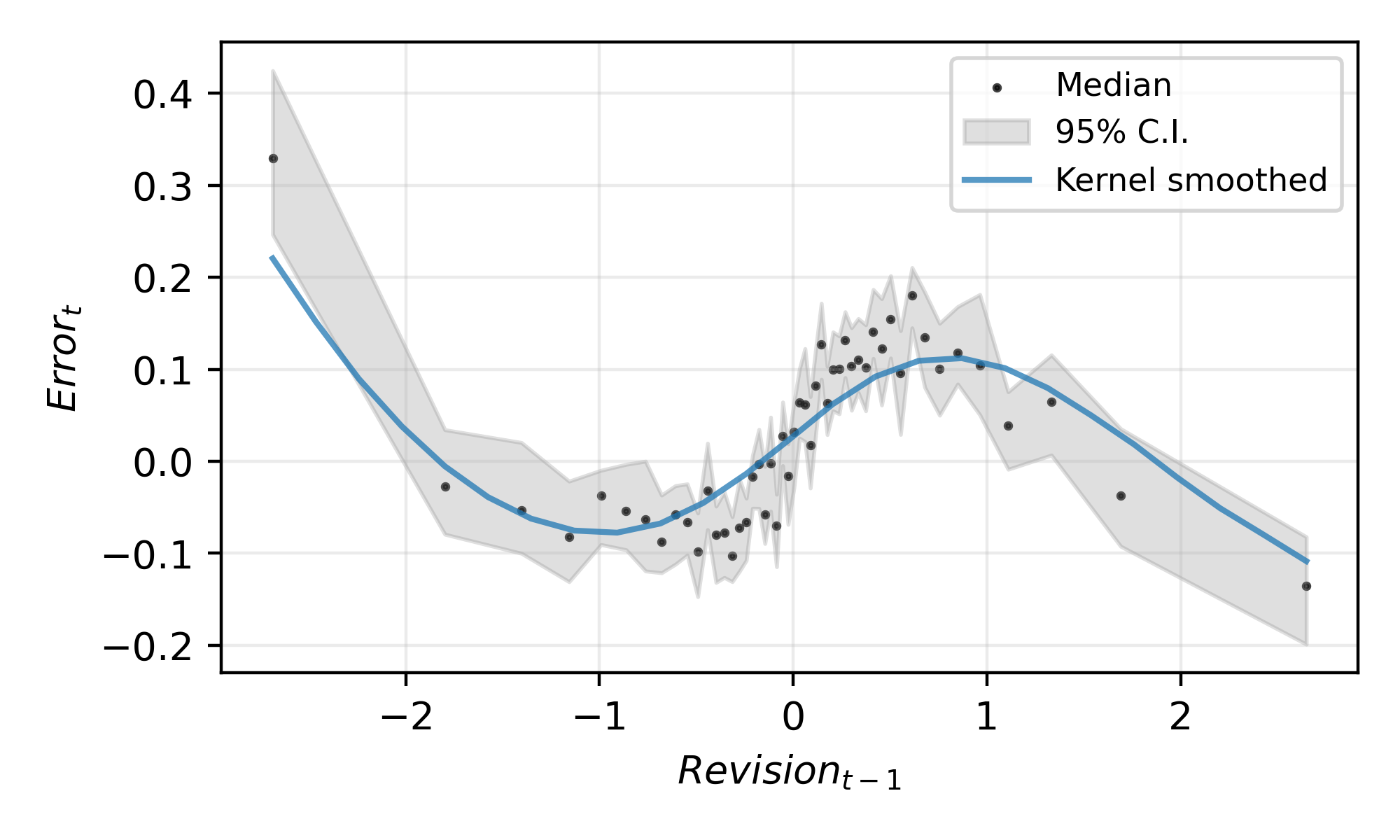}
    \end{center}
    \footnotesize Note: In this figure we use our international sample of firm revenue expectations to report the binned scatter plot of future log forecast errors $\frac{g_{t+1}-F_t g_{t+1}}{MAD_i}$ as a function of past revision $\frac{F_t g_{t+1} - F_{t-1} g_{t+1}}{MAD_i}$. The blue line is a local polynomial approximation, centered in the middle of each centile.
\end{figure}

Figure \ref{fig:zigzag_sales_norm} shows that normalized error and normalized revisions follow the same relationship as in our headline Figure \ref{fig:zigzag_sales}. In this Figure, we simply show the binned scatter plot of future log forecast errors $\frac{g_{t+1}-F_t g_{t+1}}{MAD_i}$ as a function of past revision $\frac{F_t g_{t+1} - F_{t-1} g_{t+1}}{MAD_i}$. This suggests that the non-linear relationship does not stem from firm volatility heterogeneity.

Another natural prediction of our forecasting model is that current and past forecast errors should follow a similar relationship. This comes from the fact that in our linear forecasting model, errors and revisions are proportional (equation \ref{ERRREVlin}). Thus, it mechanically follows that if error and lag revisions are linked by the S-shaped curve of Figure \ref{fig:zigzag_sales_norm}, then error and lagged error should follow the same relationship.

\begin{figure}[htbp!]
    \begin{center}
    \caption{Revenue Forecast Error as a Function of Past Error (Normalized)}
    \label{fig:zigzag_err_err_norm}
    \includegraphics{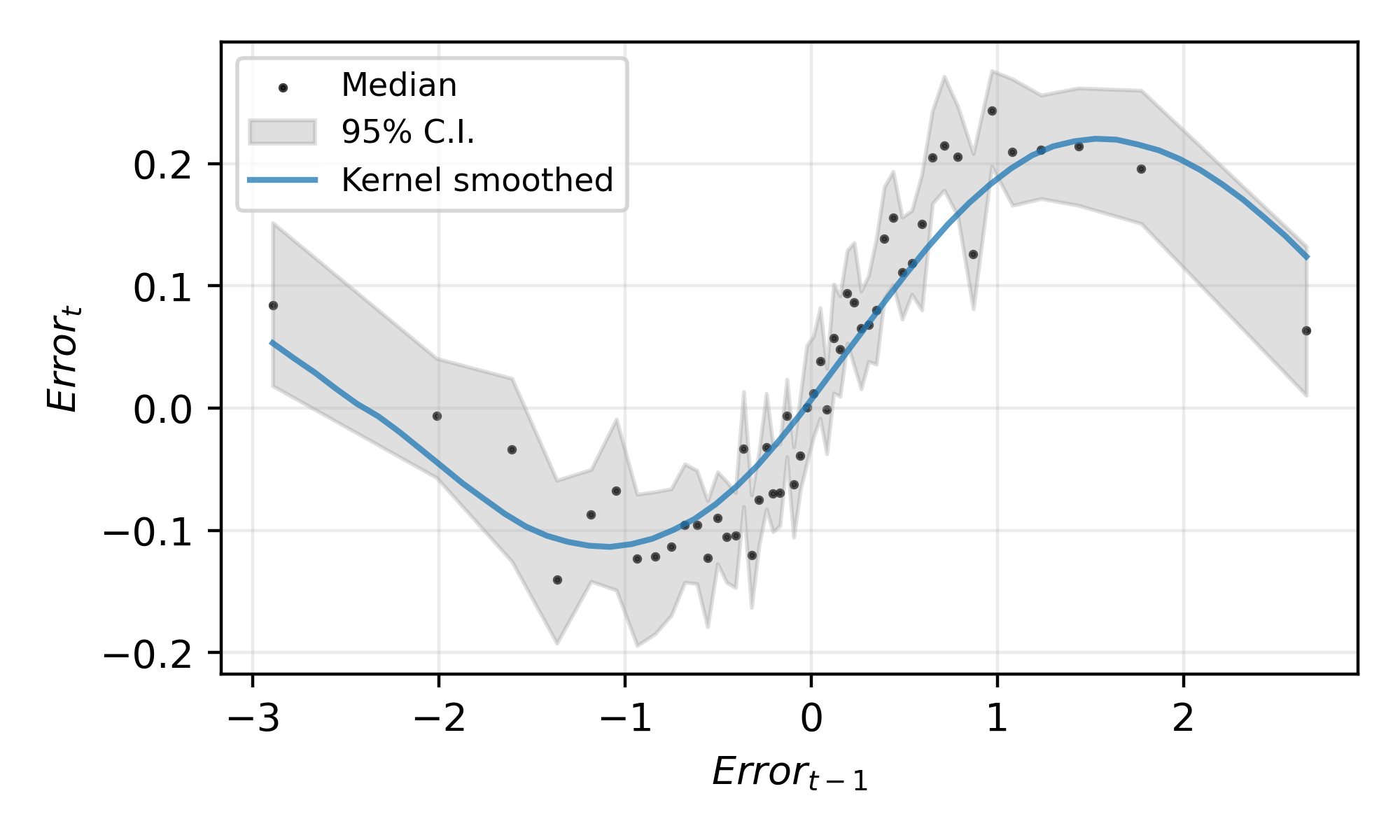}
    \end{center}
    \footnotesize Note: In this figure we use our international sample of firm revenue expectations to report the binned scatter plot of future log forecast errors $\frac{g_{t+1}-F_t g_{t+1}}{MAD_i}$ as a function of past error $\frac{g_{t} - F_{t-1} g_{t}}{MAD_i}$. The blue line is a local polynomial approximation, centered in the middle of each centile.
\end{figure}

We look at the relation between error and lagged error in Figure \ref{fig:zigzag_err_err_norm}. It shows the binned scatter plot of future log forecast errors $\frac{g_{t+1}-F_t g_{t+1}}{MAD_i}$ as a function of past error $\frac{g_{t} - F_{t-1} g_{t}}{MAD_i}$, both of them normalized by firm-level volatility. As can be seen from this figure, the forecast errors follow a linear relationship for intermediate values (until about 1 unit of volatility), but the relationship reverses for larger past errors.

\subsection{Evidence from Returns}

\begin{table}[htbp!]
  \centering
  \caption{Sample size by exchange (returns)}
  \label{tab:index_date_ret}
    \begin{tabular}{lrrrrrr}
    \toprule
    Index &  Total &  2000 &  2005 &  2010 &  2015 &  2020 \\
    \midrule
    AEX   &    5892 &      0 &    290 &    300 &    278 &    283 \\
    AS5   &   47252 &      0 &   2127 &   2340 &   2164 &   2361 \\
    CAC   &    9946 &      0 &    475 &    474 &    474 &    480 \\
    DAX   &    6531 &      0 &    360 &    360 &    360 &    351 \\
    HSC   &    5946 &      0 &      0 &    281 &    435 &    585 \\
    HSI   &    8949 &      0 &      0 &    522 &    582 &    597 \\
    IBE   &    8663 &      0 &    398 &    419 &    410 &    419 \\
    IND   &    5377 &      0 &      0 &    360 &    360 &    348 \\
    KOS   &   36595 &      0 &      0 &   2381 &   2345 &   2369 \\
    MID   &  120017 &   4585 &   4773 &   4770 &   4583 &   4750 \\
    NDX   &   21624 &      0 &   1171 &   1200 &   1234 &   1197 \\
    NIF   &    6794 &      0 &      0 &     52 &    599 &    600 \\
    NKY   &   60783 &   2642 &   2674 &   2698 &   2676 &   2698 \\
    OMX   &    7565 &      0 &    348 &    360 &    360 &    360 \\
    RAY   &  830673 &  29792 &  32936 &  33833 &  31915 &  33247 \\
    SMI   &    3715 &      0 &      0 &    233 &    240 &    228 \\
    SPT   &   10200 &      0 &      0 &    707 &    708 &    720 \\
    SX5   &    9744 &      0 &      0 &    600 &    600 &    587 \\
    TOP   &    5229 &      0 &      0 &      5 &    504 &    452 \\
    TPX   &  518734 &  16490 &  19074 &  19783 &  21722 &  25588 \\
    TWY   &   13663 &      0 &      0 &    355 &   1168 &   1028 \\
    UKX   &   29775 &   1086 &   1180 &   1196 &   1180 &   1187 \\
    \bottomrule
    \end{tabular}
\end{table}

Our model in Section \ref{returns_model} predicts that past returns should predict future returns in a non-linear way. We now provide evidence on returns based on CFM's international monthly stock returns data described in the Data section of this paper. 

\begin{figure}[htbp!]
    \caption{Binscatter Plot of Returns by Past Returns}
    \vspace{.2in}
    \begin{center}
    \footnotesize
    Panel A: Entire Sample \hspace{1.2in} Panel B: US vs International
    \includegraphics[scale=.5]{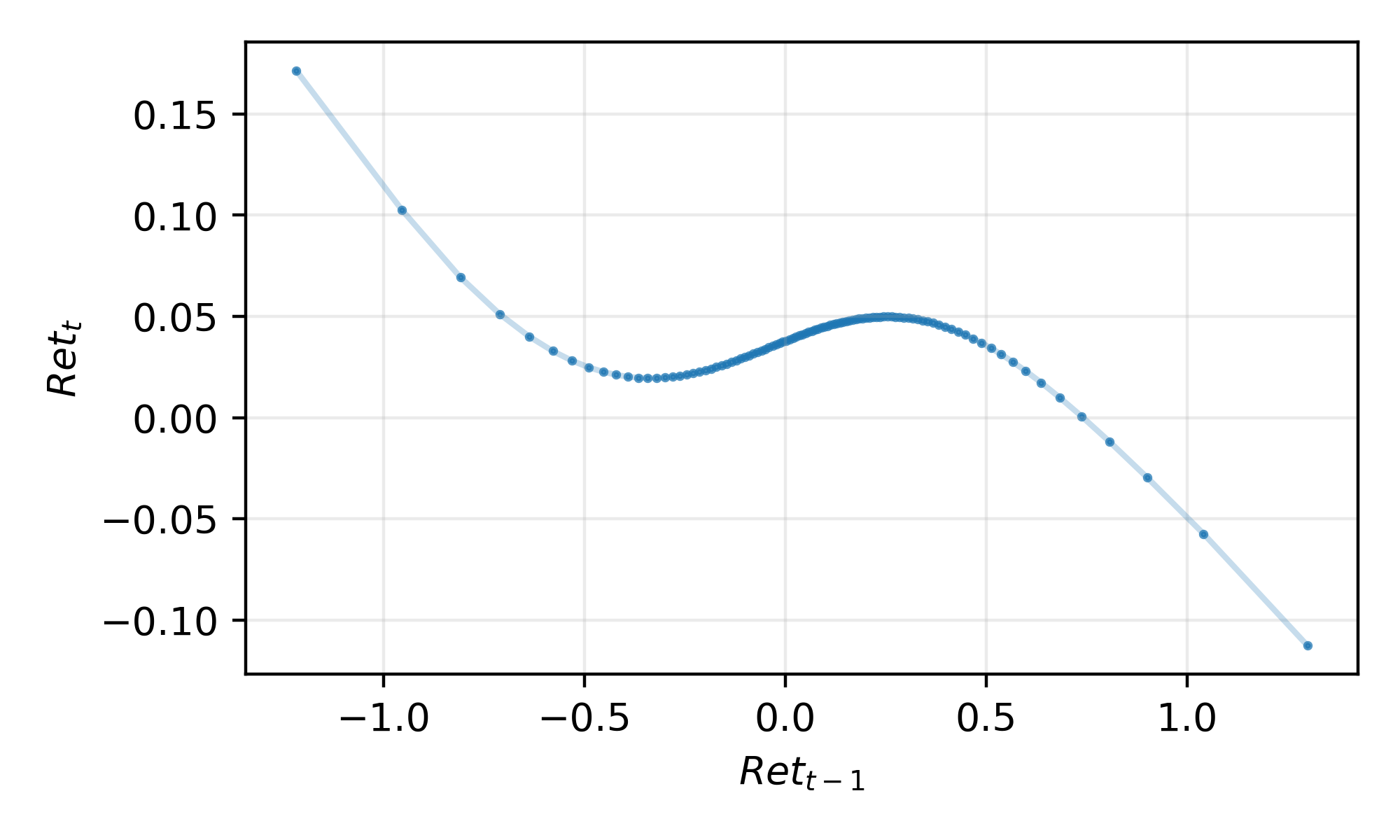}\hspace{.1in}
    \includegraphics[scale=.5]{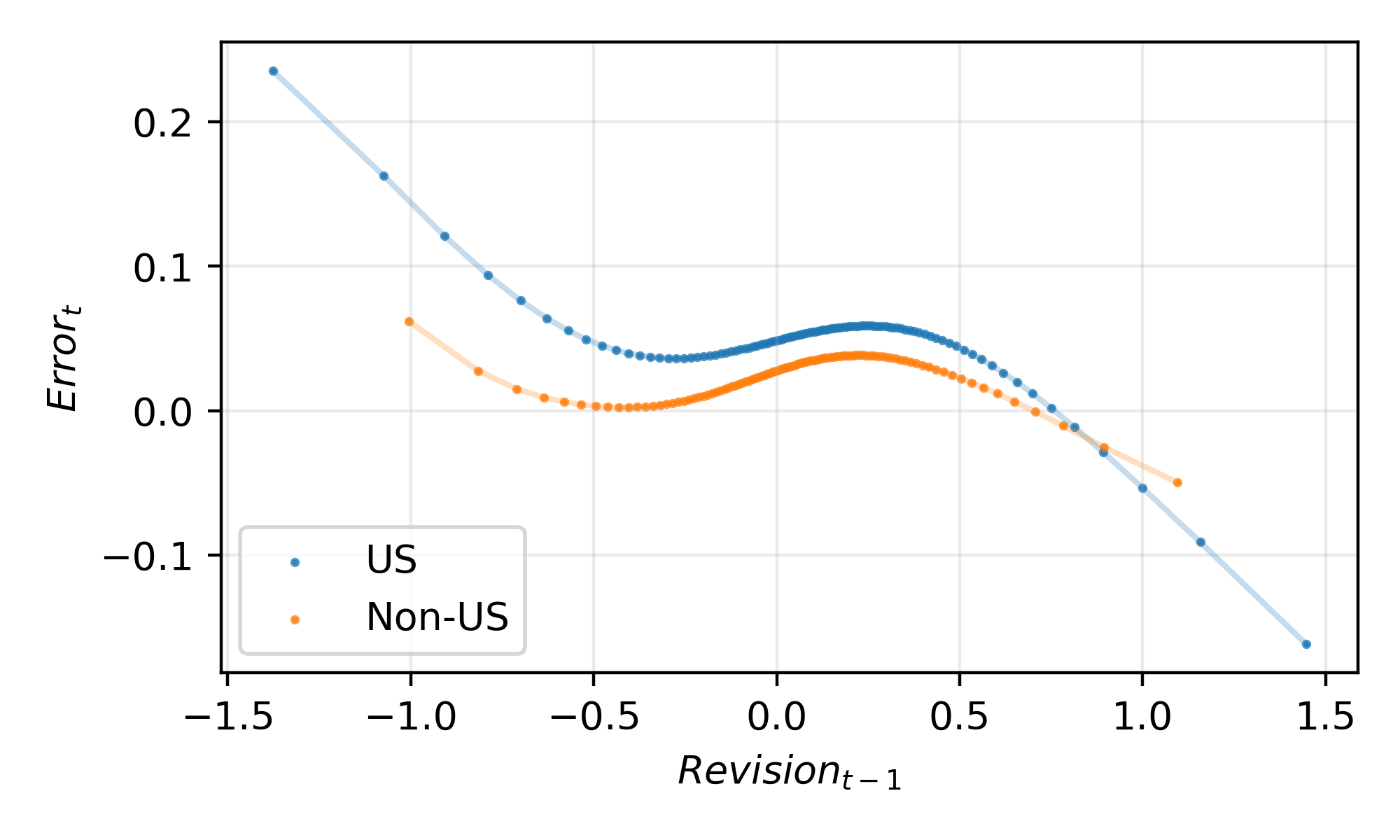} \\
    \vspace{.2in}
    Panel C: By holding Period \hspace{1.2in} Panel D: By 
    S index
    \includegraphics[scale=.5]{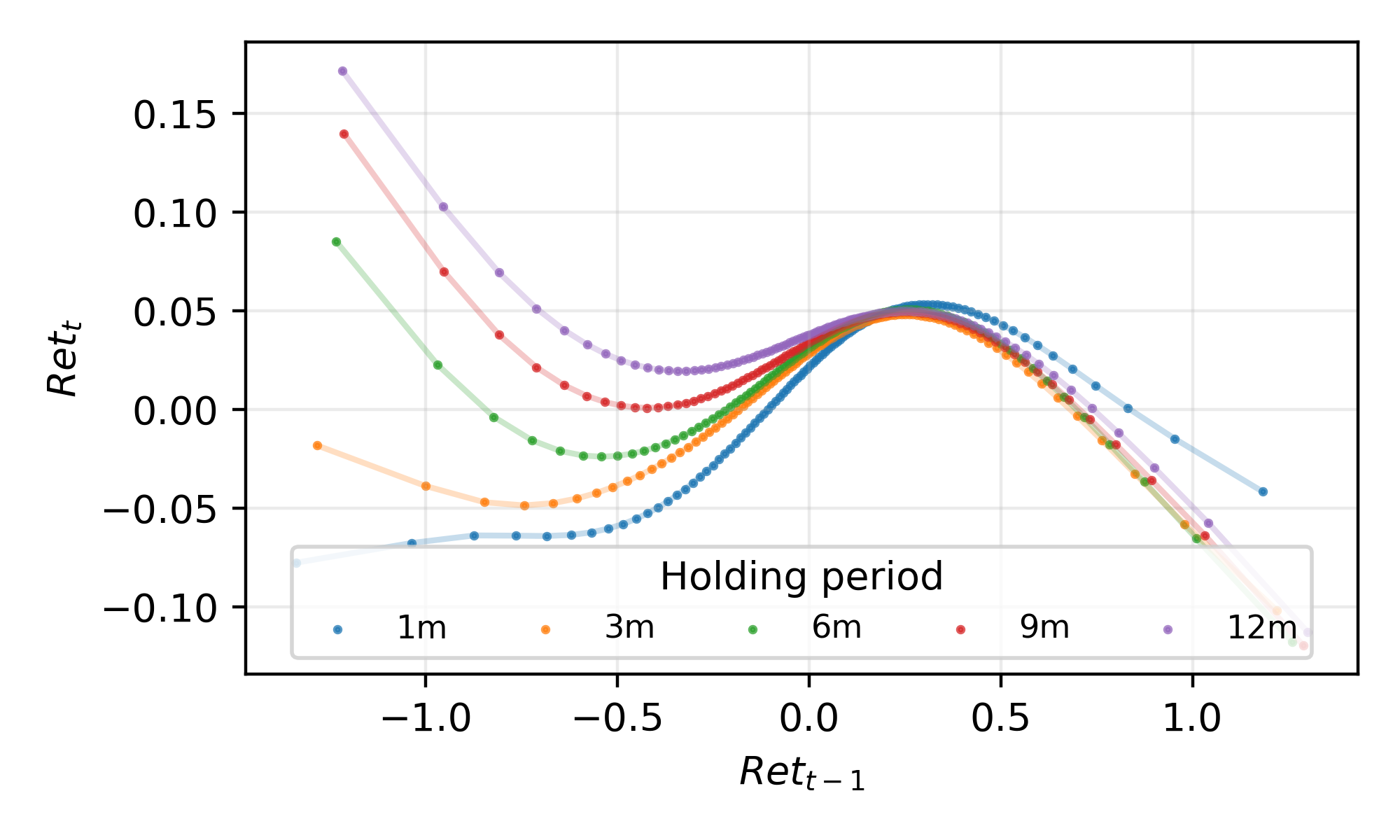}\hspace{.1in}
    \includegraphics[scale=.5]{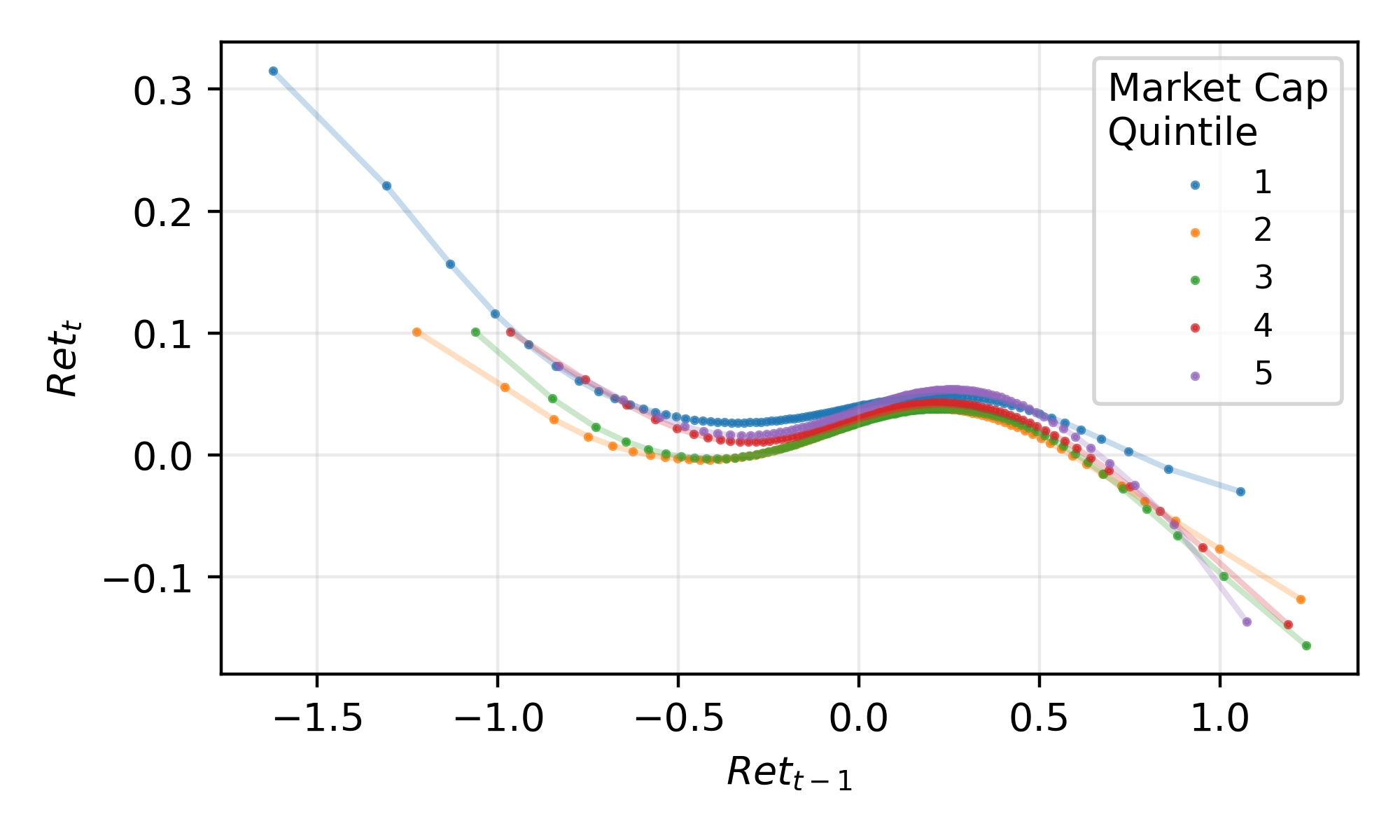}
    \label{fig:zigzag_ret}
    \end{center}
\footnotesize Note: These 4 panels represent smoothed binned scatter plots of future log returns as a function of log past cumulative returns of the past 12 months excluding the last month. Panel A is the entire sample, Panel B splits the sample into US and International stocks, Panel C computes future returns using different holding periods, and Panel D by size quintile.
    
\end{figure}

In Figure \ref{fig:zigzag_ret} we first show a smoothened binscatter plot of future returns on past returns. Future returns are monthly and past returns are calculated over the past 12 months excluding the last month of returns, as is common in the literature on stock momentum. The only difference here with the standard literature is that we take the log of returns (this is done because this analysis tends to focus on extreme past returns). 

Figure \ref{fig:zigzag_ret} shows the binned scatter plot for different splits of the data. Panel A looks at the entire dataset and provides a picture consistent with our prediction: There is momentum for most levels of past returns, but for extreme values it is mean-reversion that prevails. Panel B shows that this pattern holds both on US data and non-US returns. Panel C investigates the role of various holding periods, i.e. looking for future returns over the following 1, 3, 6, 9 and 12 months. We find that the S-shaped curve emerges as soon as this holding period is longer than one month. In panel D, we sort stocks into market cap quintiles at the index-month level. Even when examining different sizes of stocks, the S-shaped pattern is to be seen everywhere. 

This finding suggests that the performance of traditional momentum strategies could be ``boosted'' by allowing for a region of reversal in both tails. To test this, we consider a self-financing strategy that goes long on momentum for moderate values of the momentum signal, and goes short on a momentum (i.e. long on reversal) for more extreme values of the signal. Let $s_{i,t-1}$ be a momentum signal calculated from past returns. Specifically, we calculate the momentum signal as the cross-sectional rank transform of cumulative returns over 11 months from $t-12$ to $t-1$, normalized such that $s_{i,t-1}=0.5$ for firm with the greatest past returns, and $s_{i,t-1}=-0.5$ for the firm with the least. A portfolio with weights $w(s_{it})$ is then formed at time $t$ as follows: 
$$
w_{it}=\begin{cases}
			0.5 - \frac{s_{i,t-1}}{a} & \text{if $s_{i,t-1} \le a$} \\
			\frac{s_{i,t-1}-b}{b - a} - 0.5 & \text{if $a<s_{i,t-1} \le b$} \\
			0.5 - \frac{s_{i,t-1} - b}{1 - b} & \text{if $b \le s_{i,t-1}$}
		 \end{cases}
$$

where $a$ and $b$ are constant ``inflection'' points at which our strategy flips from reversal to momentum, and then from momentum to reversal. 

\begin{figure}[htbp!]
    \begin{center}
    \caption{The Error-Revision relationship: Sample Splits}
    \label{fig:zigzag_strat}
    \includegraphics[scale=.33]{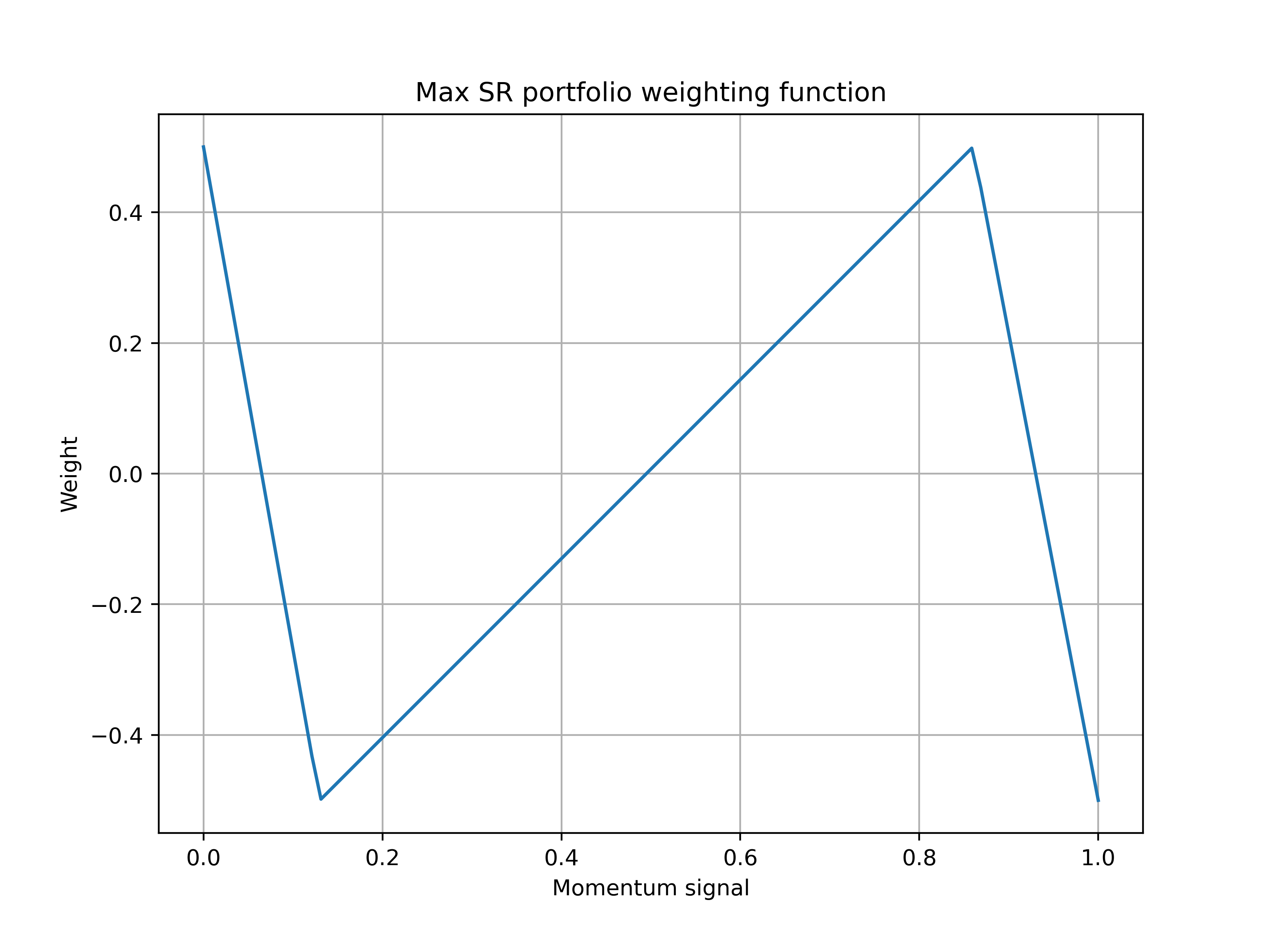} \includegraphics[scale=.33]{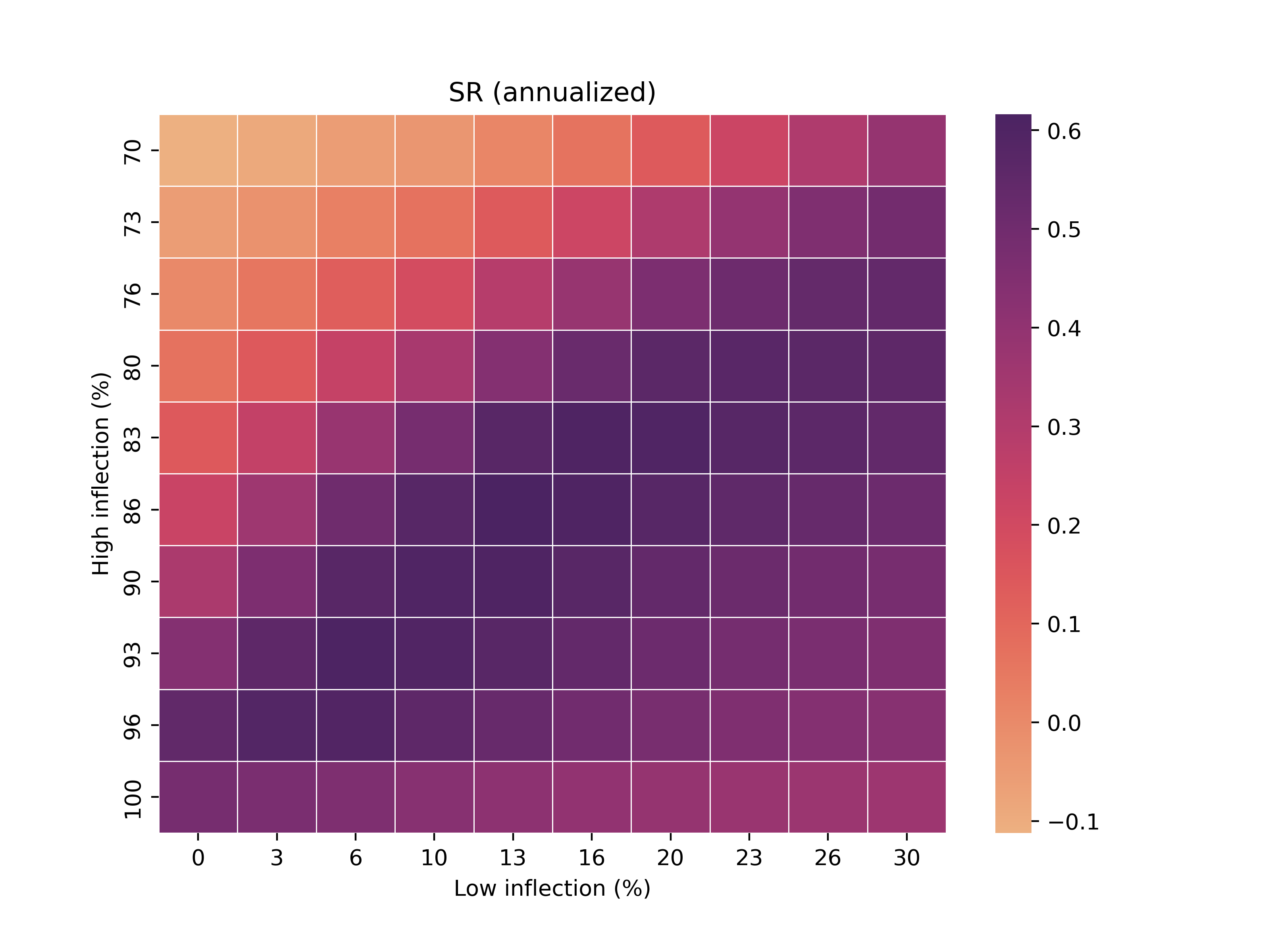} \\
    \vspace{.2cm}
    \footnotesize
    Panel A: Portfolio weights \hspace{1in} Panel B: Sharpe ratio by inflection point
    \end{center}
    \footnotesize Note: Portfolio formed across all firms in our equity sample, both U.S. and international. See text for details of momentum signal construction.
\end{figure}

Our findings are depicted in figure \ref{fig:zigzag_strat}. The inflection points of our strategy, $a$ and $b$ are chosen to optimize the Sharpe ratio of our strategy over the sample period. Panel A shows the Sharpe ratio maximizing weighting function, while panel B shows how the Sharpe ratio varies for different upper and lower inflection points. Note that a strategy with a low inflection point of 0 and upper inflection point of 100 corresponds to a traditional momentum strategy, with no tail reversal. Clearly, our approach contains some look ahead bias, as the coefficients $a$ and $b$ are estimated on the entire sample. A more systematic investigation of these returns is beyond the scope of this paper.

In line with our earlier empirical results, we find that the Sharpe ratio of our strategy is maximized when the lower inflection point is set at the 13th cross-sectional percentile of the normalized momentum signal, and the upper inflection point is set at the 86th percentile. The maximum Sharpe ratio to this strategy (0.61) is 1.27x the Sharpe ratio we observe for a pure momentum strategy (0.48). 

\section{Conclusion}
\label{conclu}

In this paper we emphasize that boundedly rational agents, when faced with fat-tailed processes, will make predictable mistakes. In order to explore such processes, we need large samples. Our empirical research here leverages the international version of IBES which gives us a large panel of sales growth forecasts. Consistently with the firm demographics literature, we find that sales growth dynamics are well described by the sum of a short-run and and long-run processes. The long-run process is a simple Gaussian, AR1 process, but the short-run process has fat tails. As a result, a simple, linear filtering rule will not be optimal. This simple model of expectations formation matches a lot of the key features of the data.

\medskip

Natural extensions of our work consists in exploring alternative forecasts data. Macro forecasts are unlikely to provide us with non-Gaussian processes and are in general too sparse to measure the tails of the DGP with enough accuracy. Within IBES studying EPS forecasts is another natural research direction, although it presents a scaling challenge. Growth cannot be computed for a large number of firms. Internal sales forecast from large companies could be another path.

\newpage
\bibliographystyle{ecta}
\bibliography{thick_bib_updated}

\end{document}